\documentclass{article}

\usepackage{arxiv}

\usepackage[utf8]{inputenc} 
\usepackage[T1]{fontenc}    
\usepackage{url}            
\usepackage{booktabs}       
\usepackage{amsfonts}       
\usepackage{nicefrac}       
\usepackage{microtype}      
\usepackage{lipsum}
\usepackage{graphicx}
\usepackage{natbib}
\usepackage{makecell}
\usepackage{array}
\usepackage{booktabs} 
\usepackage{amsmath}
\usepackage{amssymb}
\usepackage[hidelinks]{hyperref}
\usepackage{graphicx} 
\usepackage[utf8]{inputenc} 
\usepackage[strict]{changepage} 
\usepackage{float} 
\usepackage{textcomp} 
\usepackage{lmodern} 
\usepackage{caption}
\usepackage{lscape} 
\usepackage{tabularx} 

\graphicspath{ {./images/} }

\title{Color, Sentiment, and Structure: A Comparative Study of Instagram Marketing Across Economies}

\author{
 Ritesh Konka \\
  Dwarkadas J. Sanghvi College Of Engineering,\\ Mumbai, India \\
  \texttt{riteshkonkawork@gmail.com} \\
   \And
 Pranali Kurani \\
  Dwarkadas J. Sanghvi College Of Engineering,\\ Mumbai, India \\
  \texttt{pranalikurani12@gmail.com} \\
}

\begin{document}
\maketitle
\begin{abstract}
Instagram has become a key platform for global food brands to engage diverse audiences through visual storytelling. While previous research emphasizes content-based strategies, this study bridges the gap between content and context by examining how aesthetic elements—such as dominant image colors and caption sentiment—and structural factors like GDP, population, and obesity rates collectively shape consumer engagement. Using a multimodal analysis of Instagram posts from major food outlets across developed and developing countries, we assess how color schemes and sentiment influence key engagement metrics. We then extend this analysis with regression modeling to evaluate how these macroeconomic and demographic variables moderate engagement. Our results reveal that engagement patterns vary widely across regions. In developing countries, color combinations like off-white and green significantly enhance interactions, and GDP is a strong positive predictor of engagement. Conversely, in developed countries, a larger population boosts engagement while higher GDP correlates with reduced user attention. Obesity rates show a mixed influence, moderately enhancing likes in some regions while lowering comments in others. These findings highlight the critical need for localized digital marketing strategies that align content design with structural realities to optimize audience engagement.
\end{abstract}


\section{Introduction}
Social media has become a crucial tool for businesses to connect with customers \citep{HANNA2011265}, build brand identity \citep{jurivsova2020building,jokinen2016branding}, and foster long-term loyalty in today’s digital landscape \citep{julian2012using}. Among the social media platforms, Instagram stands out for its ability to support visual storytelling. With over 2 billion active users as of 2024 \citep{statistaGlobalInstagram}, Instagram provides businesses with unmatched access to a global audience, making it an essential component of digital marketing strategies. Its emphasis on visual content is especially effective in industries like fashion, lifestyle, and food \& beverage (F\&B), where aesthetics drive consumer appeal. Brands like MCD, Taco Bell, and Ben \& Jerry’s leverage Instagram to craft and communicate their unique personalities through curated visuals, user-generated content, and diverse post types \citep{ginsberg2015instabranding}.Additionally, interactive tools like polls, Q\&As, and shoppable posts facilitate two-way communication, enabling brands to better understand consumer preferences and strengthen loyalty \citep{shukla2025}

Analyzing the rich blend of content on this platform requires a sophisticated approach. Multimodal marketing refers to the integration of multiple content modes - primarily visual and textual elements - to craft more compelling and engaging brand messages on social media. In the food and beverage (F\&B) industry, Instagram is particularly powerful in this regard, as it appeals to consumer sensory needs through its highly visual and interactive platform \citep{twibiagencyFoodBeverage}. Research shows that Instagram's emphasis on visual content has a stronger influence on consumer behavior than text-based marketing, with food images driving more significant engagement. Recent advancements in multimodal analysis \citep{kim2024multi}, which integrates text and images, offers deeper insights into user engagement by analyzing both textual and visual elements in social media posts. This method is particularly effective at identifying patterns that single-mode analysis often overlooks. For example, \citep{PHILP2022736} finds that visually appealing food images that follow familiar presentation styles consistently
generate higher interaction. Combining text and image features improves predictions of social media popularity \citep{WANG2023101490}, while multimodal approaches incorporating contextual and social data achieve great accuracy in predicting engagement metrics like photo view counts \citep{8396998}. Advances in deep learning, such as self-attention mechanisms \citep{9656132,zhang2024improving}, have further enhanced predictive models by fusing semantic and numeric data, improving accuracy when analyzing content-related posts over short time frames. These developments highlight the growing sophistication and effectiveness of multimodal analysis in understanding and predicting social media engagement across various platforms.

Despite notable advancements in understanding social media engagement, a significant gap remains in how marketing strategies are adapted to different geographic and structural contexts—particularly in the food and beverage (F\&B) sector. While numerous studies have examined content-based strategies such as image quality, sentiment, and hashtags, few link these aesthetic elements to macro-level factors like economic development, health, or population dynamics. Most existing research tends to focus on either developed or developing countries in isolation  \citep{wahid2022factors,ferrara2014online}, limiting our understanding of how multinational brands navigate cross-regional variation. Even fewer studies explore how demographic and economic indicators—such as population size, GDP, or obesity rates—shape consumer engagement on platforms like Instagram. Studies highlight the critical role of structural and economic factors in shaping social media marketing, where macroeconomic conditions (e.g., digital infrastructure, market maturity, competitive intensity) influence firms' strategies across countries, while health and demographic indicators like obesity rates and population size drive engagement patterns, with higher obesity rates correlating to greater followership of unhealthy food and beverage brands \cite{li2024understanding,gu2021associations}.

This study bridges that gap by analyzing Instagram marketing strategies used by major multinational food chains—including KFC, Subway, Domino’s, and MCD—across both developed and developing countries. Specifically, we adopt a two-level analytical approach: first, we perform a multimodal analysis combining visual elements (dominant color schemes) and textual elements (sentiment in post captions) to assess their influence on engagement metrics. Second, we extend our analysis with a regression model to evaluate how macroeconomic (GDP, population) and health-related (obesity rate) factors interact with engagement outcomes. By doing so, we investigate how visual and emotional marketing strategies interface with broader structural realities. This leads us to our central research question: How do content aesthetics and structural contexts together shape engagement outcomes for global food brands on Instagram?

\section{Related Work}
\subsection{Instagram Marketing and Regional Strategy Gaps}

In recent years, Instagram has emerged as a crucial platform for digital marketing, providing businesses with powerful tools to connect with a global audience through visually engaging content. Its features—such as Stories, Reels, and shoppable posts—enable brands to craft compelling narratives and foster interactive relationships with consumers. Research shows that discounts and influencer endorsements on social media significantly influence purchase decisions, particularly for personal care and hygiene products \citep{Ananthsai2023-ey}, highlighting the importance of strategic content creation. Furthermore, Instagram enhances visibility for small businesses and contributes to stronger advertising impact through features like Reels \citep{inbook}.

In the food and beverage (F\&B) sector specifically, Instagram has proven to be a valuable platform for promoting healthier diets, with visually appealing posts positively affecting consumer purchase intentions \citep{fernandes2018instagram}. However, concerns about the promotion of unhealthy products via influencer marketing underscore the need for more regulatory oversight \citep{Reagan2020adOI}. Instagram’s broad utility has led to the development of strategic frameworks that align platform-based marketing with brand missions, though its fragmented and context-sensitive nature presents challenges \citep{felix2017elements}.

Despite the rich body of literature on Instagram marketing, most existing studies tend to analyze content strategies within either developed or developing countries—rarely both. Furthermore, few link these content strategies to broader contextual factors like national wealth, population size, or public health indicators. As a result, limited insight exists into how marketing approaches should be adapted to regional or socioeconomic realities. Our study addresses this limitation by comparing engagement strategies and outcomes across economic contexts, particularly between developed and developing countries.

\subsection{Multimodal Analysis in Digital Engagement Research}

Alongside the growth of platform-specific marketing research, multimodal analysis has gained prominence as a method for understanding online user behavior. Multimodal approaches integrate text, images, and other media types to gain deeper insights into how content is perceived and engaged with on digital platforms \citep{Liu2023-di}. In the context of Instagram, studies have shown that visually familiar or aesthetically appealing food images tend to attract higher engagement, reaffirming the platform’s visual power \citep{PHILP2022736}. Moreover, combining image data with textual features such as sentiment or entertainment value significantly enhances the prediction of user interactions \citep{10.1145/2964284.2967210, crowe2018predicting}. This integration also improves the accuracy of sentiment classification systems, offering richer insights into consumer opinions \citep{soton360546}.

Beyond marketing, multimodal analysis has been applied to domains such as mental health, where models combining visual, textual, and behavioral data help detect depression and suicidal ideation with high accuracy \citep{yazdavar2020multimodal, ramirez2020detection, malhotra2020multimodal}. Similar frameworks have been used to identify early signs of gang violence \citep{blandfort2019multimodal}, improve humanitarian response during disasters \citep{ofli2020analysis, gautam2019multimodal}, and even detect underage drinking patterns on social media platforms \citep{pang2015monitoring}. Recent datasets like the Love-Hate Dataset \citep{10.1145/3589335.3651966} have further explored emotional polarity in multimodal content during conflict situations, highlighting patterns of engagement across different sentiments.

In marketing research, multimodal techniques have proven useful for generating sentiment-informed captions, detecting marketing intent, and enhancing audience targeting \citep{zhang2021multimodal, harzig2018multimodal}. In sectors like automotive and beauty advertising, visual and textual cues have been shown to strongly influence consumer perception \citep{Kuswandini2018/07, amatullah2019analysis}. Additionally, integrated multimodal strategies in platforms like YouTube have improved the measurement and understanding of ad engagement \citep{vedula2017multimodal}. Expanding the geographic dimension, studies such as the Social Cyber Geography Inventory \citep{ng2025socialcybergeographicalworldwide} demonstrate how automated engagement patterns vary across languages and regions in global digital spaces.

However, a key limitation in the current literature is the lack of research exploring how multimodal elements interact with regional economic, cultural, or demographic contexts to shape consumer engagement. While these methods have demonstrated strong predictive capabilities, they are rarely applied to cross-country comparative studies—especially those combining content features with macroeconomic indicators. Our study addresses this gap by conducting a comprehensive multimodal analysis of Instagram content from major global food brands. Specifically, we analyze how textual features (sentiment, hashtags) and visual features (dominant color schemes) influence engagement differently across developed and developing countries. We further extend this framework by introducing a regression analysis of structural factors—GDP, population, and obesity rates—thereby offering a more nuanced, context-aware understanding of engagement behavior on social media.

\section{Data Collection}\label{sec:data_collection}
We leveraged the CrowdTangle API to extract comprehensive data from the official Instagram accounts of seven leading food and beverage brands: Starbucks, Burger King, MCD, Domino’s, Pizza Hut, KFC, and Subway. The dataset includes detailed post-level information, capturing key attributes and their descriptions, as outlined in \autoref{tab:instagram_post_data_attributes} in Appendix A. These attributes include engagement metrics (likes, comments, views, total interactions) and content features (captions, post created date, posting time), enabling a thorough analysis of their social media strategies and audience interactions.

To uncover patterns and strategies that drive higher engagement on Instagram, we categorized brand posts by their country of origin and analyzed a range of engagement metrics. Our study includes both developed and developing countries, enabling a comparative analysis of marketing techniques and their effectiveness across diverse economic contexts. The countries listed in \autoref{tab:developed_and_developing_countries} were selected based on their development status as per GDP, the presence of the majority of the food outlets in those countries, and the availability of data via the CrowdTangle API.

The classification of countries as 'developed' or 'developing' was primarily based on the criteria set forth by major international bodies, including the United Nations \citep{UN_WESP_2025_MidYear}, the World Bank \citep{worldbankWorldBank}, and the International Monetary Fund \citep{imfWorldEconomic}. However, our classification also considered marketing dynamics, leading to one notable exception. While South Korea is recognized as a high-income developed economy by these institutions \citep{koreaAgencyUpgrades}, it was grouped with developing countries for this analysis. This decision reflects its status as a highly dynamic 'emerging market' in influential business frameworks such as the MSCI Emerging Markets Index \citep{MSCI_EM_Factsheet_2025} and is justified by its unique consumer landscape, characterized by rapid trend adoption and digital engagement patterns that align more closely with other fast-growing markets in our sample.

\begin{table}[ht] 
\centering 
\renewcommand{\arraystretch}{1.5} 
\caption{List of Developed and Developing Countries in Our Study}
\label{tab:developed_and_developing_countries}
\small\sf 
\resizebox{0.5\textwidth}{!}{ 
\begin{tabular}{ll} 
\toprule 
\textbf{Developed Countries} & \textbf{Developing Countries} \\
\midrule 
Spain & India \\ \hline
France & Brazil \\ \hline
Italy & Mexico \\ \hline
Canada & Saudi Arabia \\ \hline
Australia & Egypt \\ \hline
Japan & Turkey \\ \hline
United Kingdom & Malaysia \\ \hline
Ireland & South Korea \\ \hline
Netherlands & UAE \\ \hline
Germany & Indonesia \\ \hline
& Argentina \\ 
\bottomrule 
\end{tabular}}
\end{table}

\section{Methodology}
Our methodology consists of two main components, multimodal analysis and regression analysis, each designed to uncover different dimensions of user engagement. The multimodal analysis focuses on content-driven engagement by examining features within the posts themselves, while the regression analysis explores context-driven engagement by incorporating external, country-level factors.

In the first stage, we use multimodal analysis to understand how content-level features influence user interactions. This involves three steps: first, we employ descriptive statistics to summarize key engagement metrics (e.g., likes, comments). Second, our text analysis involves conducting sentiment analysis on post captions to evaluate emotional tone and extracting frequently used hashtags to identify content trends. Third, the image analysis focuses on detecting dominant colors in post visuals to assess aesthetic patterns.

Building on this, the second stage introduces a regression analysis of structural factors to examine how macro-level variables—GDP, national population, and adult obesity rates—influence engagement across regions. This component allows us to explore how socioeconomic and health-related contexts shape digital behavior beyond the content itself.

\begin{figure}[htp]
\centering
\includegraphics[width=1\textwidth]{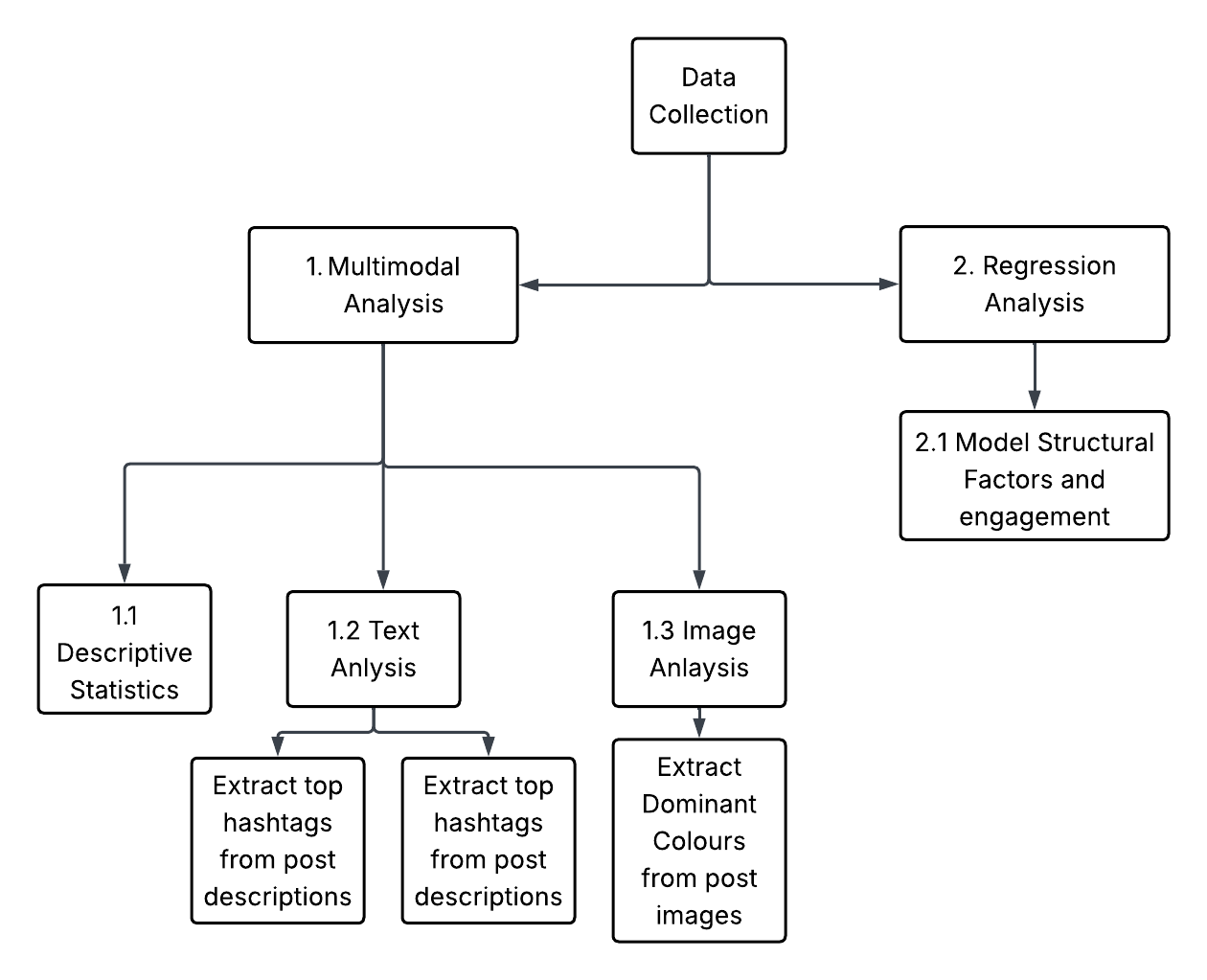}
\caption{Flowchart of methodology employed in this study.}
\label{fig:methodology_flowchart}
\end{figure}

These analytical stages are visually summarized in the flowchart shown in \autoref{fig:methodology_flowchart}, which outlines the progression from data collection to both multimodal and structural analyses.

\subsection{Multimodal Analysis}

\subsubsection{Descriptive Statistics}
We summarize key engagement metrics, including comments, likes, views, followers, total interactions and overperformance scores (how well the total interactions of a post performed relative
to the average total interactions of the account for that type of post) for all food outlets in both developed and developing countries. This analysis, presented in \autoref{tab:descriptive_stats_developed} and \autoref{tab:descriptive_stats_developing}, reveals patterns of consumer interaction and highlights how brand engagement varies in different economic contexts.

By calculating the average and standard deviation of these metrics, we identified general trends in user engagement, enabling a comparative view between developed and developing regions. This comparison underscores differences in audience behavior and brand resonance, offering preliminary insights into how certain outlets may perform differently depending on their geographic and economic environment.

\textbf{1. Developed Countries :} 

\begin{table*}[]
    \centering
    \caption{Descriptive statistics of developed countries}
    \label{tab:descriptive_stats_developed}
    \footnotesize
    \setlength{\tabcolsep}{4pt} 
    \begin{tabular}{lccccccc}
        \toprule
        \textbf{Metric} & \textbf{Starbucks} & \textbf{MCD} & \textbf{PizzaHut} & \textbf{BurgerKing} & \textbf{Dominos} & \textbf{Subway} & \textbf{KFC} \\
        \midrule
        \makecell[l]{Comments} & \makecell{50.00 \\ $\pm$ 104.44} & \makecell{226.63 \\ $\pm$ 1951.16} & \makecell{123.30 \\ $\pm$ 249.70} & \makecell{28.93 \\ $\pm$ 125.52} & \makecell{56.03 \\ $\pm$ 457.87} & \makecell{205.45 \\ $\pm$ 2205.42} & \makecell{25.15 \\ $\pm$ 165.30} \\
        \addlinespace
        \makecell[l]{Likes} & \makecell{1857.8 \\ $\pm$ 3301.02} & \makecell{3512.20 \\ $\pm$ 9339.17} & \makecell{38673.27 \\ $\pm$ 94874.70} & \makecell{739.59 \\ $\pm$ 1658.18} & \makecell{2078.69 \\ $\pm$ 4498.58} & \makecell{810.69 \\ $\pm$ 2250.10} & \makecell{956.19 \\ $\pm$ 2659.12} \\
        \addlinespace
        \makecell[l]{Followers} & \makecell{239360.97 \\ $\pm$ 189690.19} & \makecell{208006.21 \\ $\pm$ 109318.20} & \makecell{357204.57 \\ $\pm$ 438511.60} & \makecell{23457.85 \\ $\pm$ 15100.63} & \makecell{124658.18 \\ $\pm$ 130706.01} & \makecell{88515.17 \\ $\pm$ 60427.79} & \makecell{37945.73 \\ $\pm$ 20903.78} \\
        \addlinespace
        \makecell[l]{Views} & \makecell{48.50 \\ $\pm$ 605.68} & \makecell{348.41 \\ $\pm$ 3392.42} & \makecell{368.85 \\ $\pm$ 6074.96} & \makecell{3.96 \\ $\pm$ 65.84} & \makecell{87.30 \\ $\pm$ 1068.54} & \makecell{0.0 \\ $\pm$ 0.0} & \makecell{2.09 \\ $\pm$ 34.51} \\
        \addlinespace
        \makecell[l]{Total \\ Interactions} & \makecell{1907.80 \\ $\pm$ 3354.97} & \makecell{3738.83 \\ $\pm$ 10092.91} & \makecell{38796.58 \\ $\pm$ 95045.89} & \makecell{768.52 \\ $\pm$ 1683.58} & \makecell{2134.72 \\ $\pm$ 4573.44} & \makecell{1016.14 \\ $\pm$ 3508.01} & \makecell{981.34 \\ $\pm$ 2695.31} \\
        \addlinespace
        \makecell[l]{Overperforming \\ Score} & \makecell{-8.19 \\ $\pm$ 62.44} & \makecell{0.09 \\ $\pm$ 3.25} & \makecell{0.50 \\ $\pm$ 12.14} & \makecell{-0.72 \\ $\pm$ 10.86} & \makecell{-2.78 \\ $\pm$ 151.57} & \makecell{-110.68 \\ $\pm$ 498.13} & \makecell{1.81 \\ $\pm$ 21.79} \\
        \bottomrule
    \end{tabular}
    \par
    {* Metrics represent mean $\pm$ standard deviation for posts from each outlet.}
\end{table*}

In developed countries, engagement metrics show that KFC is a dominant outlier in visibility, averaging a remarkable $38{,}673.27\ \pm\ 94{,}874.70$ likes and $38{,}796.58\ \pm\ 95{,}045.89$ total interactions per post, far surpassing all other brands. While KFC leads in raw interactions, MCD and Dominos generate the highest number of comments, with averages of $226.63\ \pm\ 1951.16$ and $205.45\ \pm\ 2205.42$ respectively; however, the large standard deviations indicate extreme variability in their post performance. Follower counts are highest for KFC ($357{,}204.57\ \pm\ 438{,}511.60$) and Starbucks ($239{,}360.97\ \pm\ 189{,}690.19$). The overperforming scores highlight these varied strategies, with KFC showing strong positive performance ($0.50\ \pm\ 12.14$) while Dominos significantly underperforms ($-110.68\ \pm\ 498.13$) relative to its own account average. This pattern suggests that while some brands excel at driving massive like counts, others foster more conversational engagement.

\textbf{2. Developing Countries :}

\begin{table*}[]
    \centering
    \caption{Descriptive statistics of developing countries}
    \label{tab:descriptive_stats_developing}
    \footnotesize 
    \setlength{\tabcolsep}{4pt} 
    \begin{tabular}{lccccccc}
        \toprule
        \textbf{Metric} & \textbf{Starbucks} & \textbf{MCD} & \textbf{KFC} & \textbf{PizzaHut} & \textbf{BurgerKing} & \textbf{Dominos} & \textbf{Subway} \\
        \midrule
        Comments & \makecell{90.18 \\ $\pm$ 245.81} & \makecell{196.94 \\ $\pm$ 1344.41} & \makecell{144.20 \\ $\pm$ 554.88} & \makecell{70.54 \\ $\pm$ 432.64} & \makecell{91.50 \\ $\pm$ 490.31} & \makecell{114.99 \\ $\pm$ 264.03} & \makecell{10.36 \\ $\pm$ 57.82} \\
        \addlinespace
        Likes & \makecell{4274.12 \\ $\pm$ 7013.80} & \makecell{2913.74 \\ $\pm$ 7716.53} & \makecell{2226.29 \\ $\pm$ 7293.99} & \makecell{1205.09 \\ $\pm$ 5812.97} & \makecell{2390.74 \\ $\pm$ 8689.00} & \makecell{524.18 \\ $\pm$ 972.54} & \makecell{356.20 \\ $\pm$ 1768.81} \\
        \addlinespace
        Followers & \makecell{1004390.02 \\ $\pm$ 543412.99} & \makecell{1423074.60 \\ $\pm$ 1277295.12} & \makecell{1067326.55 \\ $\pm$ 1006175.84} & \makecell{703832.15 \\ $\pm$ 775191.36} & \makecell{452697.82 \\ $\pm$ 724264.19} & \makecell{283243.63 \\ $\pm$ 193812.39} & \makecell{59710.03 \\ $\pm$ 52601.40} \\
        \addlinespace
        Views & \makecell{579.21 \\ $\pm$ 6365.35} & \makecell{332.07 \\ $\pm$ 5072.17} & \makecell{440.97 \\ $\pm$ 3202.73} & \makecell{365.75 \\ $\pm$ 6272.36} & \makecell{268.61 \\ $\pm$ 2562.94} & \makecell{22.76 \\ $\pm$ 465.84} & \makecell{1.03 \\ $\pm$ 28.98} \\
        \addlinespace
        \makecell[l]{Total \\ Interactions} & \makecell{4364.30 \\ $\pm$ 7190.40} & \makecell{3110.68 \\ $\pm$ 8512.97} & \makecell{2370.50 \\ $\pm$ 7475.19} & \makecell{1275.63 \\ $\pm$ 5889.59} & \makecell{2482.25 \\ $\pm$ 9033.32} & \makecell{639.18 \\ $\pm$ 1107.23} & \makecell{366.57 \\ $\pm$ 1786.39} \\
        \addlinespace
        \makecell[l]{Overperforming \\ Score} & \makecell{-0.13 \\ $\pm$ 3.80} & \makecell{0.21 \\ $\pm$ 7.58} & \makecell{0.16 \\ $\pm$ 5.79} & \makecell{1.69 \\ $\pm$ 19.98} & \makecell{0.15 \\ $\pm$ 9.34} & \makecell{-0.23 \\ $\pm$ 5.89} & \makecell{1.19 \\ $\pm$ 18.75} \\
        \bottomrule
    \end{tabular}
    \par
    * Metrics represent mean $\pm$ standard deviation for posts from each outlet.
\end{table*}

In developing countries, the engagement landscape shifts significantly, with Starbucks and MCD emerging as the leaders in audience interaction. Starbucks records the highest average likes ($4{,}274.12\ \pm\ 7{,}013.80$) and total interactions ($4{,}364.30\ \pm\ 7{,}190.40$), with MCD also showing strong performance in both categories. MCD, in turn, garners the highest average number of comments at $196.94\ \pm\ 1344.41$. A key distinction in this region is the massive follower counts, with MCD ($1{,}423{,}074.60\ \pm\ 1{,}277{,}295.12$), KFC ($1{,}067{,}326.55\ \pm\ 1{,}006{,}175.84$), and Starbucks ($1{,}004{,}390.02\ \pm\ 543{,}412.99$) commanding audiences that are several times larger than their counterparts in developed nations. Unlike in developed markets, the overperforming scores are more modest, with PizzaHut ($1.69\ \pm\ 19.98$) and Subway ($1.19\ \pm\ 18.75$) showing the most consistent positive performance relative to their averages. These findings indicate a different set of brand leaders and a distinct dynamic of user engagement in developing regions.

\subsubsection{\textbf{Text Analysis}}
In this study, text analysis techniques were applied to social media content from seven major food outlets, with a focus on sentiment and hashtag analysis. By examining large volumes of user-generated content, text analysis offers valuable insights into consumer perceptions, brand engagement, and emerging trends. This approach enables marketers to better understand audience sentiment and evaluate the effectiveness of their campaigns.

\paragraph{\textbf{Sentiment Analysis}}
We conducted sentiment analysis on Instagram post descriptions using a pre-trained model from Hugging Face\footnote{https://huggingface.co/lxyuan/distilbert-base-multilingual-cased-sentiments-student}. The model categorized the posts as Positive, Neutral, or Negative. This analysis was performed across posts from both developed and developing countries, enabling us to compare consumer reactions to the brands' marketing efforts in different economic contexts. By assessing the sentiment of these post descriptions, we gained insights into how effectively the emotional tone of marketing messages resonates with the target audience in different regions.

\textbf{1. Developed Countries: }

\begin{table}[h]
\caption{Sentiment analysis of fast-food brands in developed countries}\label{tab:sentiment_analysis_developed}
\centering
\renewcommand{\arraystretch}{1.2} 
\setlength{\tabcolsep}{5pt}      

\footnotesize 
\begin{tabular}{@{}lccccccc@{}}
\toprule
\textbf{Sentiment} & \textbf{Starbucks} & \textbf{KFC} & \textbf{MCD} & \textbf{Subway} & \textbf{Dominos} & \textbf{Pizza Hut} & \textbf{Burger King} \\
\midrule
Positive & 672 & 476 & 499 & 424 & 549 & 663 & 525 \\
\hline
Negative & 73 & 154 & 145 & 104 & 113 & 76 & 223 \\
\hline
Neutral  & 9 & 27 & 32 & 11 & 17 & 9 & 20 \\
\bottomrule
\end{tabular}
\end{table}

The sentiment analysis in developed countries, as shown in \autoref{tab:sentiment_analysis_developed}, reveals that positive sentiment overwhelmingly dominates across all brands, with Starbucks (672) and Pizza Hut (663) receiving the highest number of positive posts. However, a notable amount of negative sentiment exists for certain brands, particularly Burger King (223 negative mentions), followed by KFC (154) and MCD (145). Neutral sentiment is minimal across the board, indicating that consumers generally express a clear positive or negative opinion about these brands. 

\vspace{0.5cm}
\textbf{2. Developing Countries: }

\begin{table}[h]
\caption{Sentiment analysis of fast-food brands in developing countries}
\centering
\renewcommand{\arraystretch}{1.2} 
\setlength{\tabcolsep}{5pt}      

\footnotesize 
\label{tab:sentiment_analysis_developing}
\begin{tabular}{@{}lccccccc@{}}
\toprule
\textbf{Sentiment} & \textbf{Starbucks} & \textbf{KFC} & \textbf{MCD} & \textbf{Subway} & \textbf{Dominos} & \textbf{Pizza Hut} & \textbf{Burger King} \\
\midrule
Positive & 1611 & 738 & 1729 & 665 & 727 & 1092 & 1390 \\
\hline
Negative & 90 & 135 & 287 & 72 & 140 & 68 & 201 \\
\hline
Neutral  & 8 & 16 & 41 & 13 & 17 & 6 & 11 \\
\bottomrule
\end{tabular}
\end{table}

The sentiment analysis in developing countries, as shown in \autoref{tab:sentiment_analysis_developing}, indicates that MCD (1,729) and Starbucks (1,611) lead in positive sentiment, reflecting strong brand presence and reputation. However, MCD also has the highest volume of negative sentiment by a large margin (287 mentions), followed by Burger King (201). Similar to developed countries, neutral sentiment remains low, demonstrating that consumers in these regions also tend to have definitive opinions about these brands.

\paragraph{\textbf{Hashtag Analysis}}
Hashtag analysis was utilized to explore the thematic elements and trends in the social media discourse surrounding the brand. Hashtags are commonly used in social media marketing to categorize content, amplify reach, and increase visibility. By analyzing the most frequently used and co-occurring hashtags, we identified key themes and trends that reflect the brand's online presence and consumer interests.

\textbf{1. Developed Countries : }

\begin{table}[h]
\caption{Top hashtags for fast-food brands in developed countries}
\centering
\setlength{\tabcolsep}{5pt}      
\label{tab:hashtags_developed}
\begin{tabular}{@{}ll@{}}
\toprule
\textbf{Outlet} & \textbf{Hashtags} \\
\midrule
Starbucks & \makecell[l]{\#InMyStarbucksEra \#EdSheeranXStarbucks \#AutumnVariations\\ 
\#PumpkinSpiceLatte \#Starbucks} \\
\hline
Subway & \makecell[l]{\#SubMelts \#Subway \#SubwayEspaña \#BetterForYou \#Footlong} \\
\hline
MCD & \makecell[l]{\#mcdonalds \#ichliebees \#Summerdays2024 \#mcdonaldsitalia \#mclovers} \\
\hline
KFC & \makecell[l]{\#KFC \#KFCItalia \#KFCItaly \#KentuckyFriedChicken \#FinalmenteKFC} \\
\hline
Domino's & \makecell[l]{\#dominosdeutschland \#dominospizza \#dominos\\ \#dominos\_nl \#veganuary} \\
\hline
Pizza Hut & \makecell[l]{\#pizzahut \#wings \#nooneoutpizzasthehut\\ \#dublin \#dundalkpizza} \\
\hline
Burger King & \makecell[l]{\#BurgerKing \#communityburger \#haveityourway\\ \#KCCxBK \#HomeOfFootball} \\
\bottomrule
\end{tabular}
\end{table}

We extracted the top hashtags for each outlet, as shown in \autoref{tab:hashtags_developed}, which reflects diverse strategies to engage audiences in mature markets. For instance, Starbucks uses \#InMyStarbucksEra and \#EdSheeranXStarbucks to connect with pop culture, while Subway's \#BetterForYou hashtag emphasizes health-conscious product offerings. Brands also use localization, such as KFC with \#KFCItalia, and community-building, like Burger King with \#communityburger and \#HomeOfFootball. These varied examples show how brands in developed regions leverage hashtags to build distinct brand personalities and target specific consumer interests beyond simple brand awareness.

\textbf{2. Developing Countries : }

\begin{table}[h]
\caption{Top hashtags for fast-food brands in developing countries}
\centering
\label{tab:hashtags_developing}
\begin{tabular}{@{}ll@{}}
\toprule
\textbf{Outlet} & \textbf{Hashtags} \\
\midrule
Starbucks & \makecell[l]{\#StarbucksMalaysia \#EsMomentoDe \#ElOtoñoSuenaBien\\ 
\#LiveYourNow \#ViveTuAhora} \\
\hline
Subway & \makecell[l]{\#MakanFreshSubway \#SubwayMY \#EatFresh\\ 
\#SubwaySetiapHari \#SubwaySoChun} \\
\hline
MCD & \makecell[l]{\#MéquiNoBBB24 \#BBB24 \#McDonaldsTürkiye\\ \#McDonaldsGibisiYok \#McDonaldsIndia} \\
\hline
KFC & \makecell[l]{\#KFCIndonesia \#SeleraOriginal \#KFCArabia\\ 
\#JagonyaKorea \#AaoLunchKarein} \\
\hline
Domino's & \makecell[l]{\#PraTodosVerem \#DominosIndia \#GününEnGüzelAnı\\ 
\#AMaiorPizzariaDoMundo \#DominosBol} \\
\hline
Pizza Hut & \makecell[l]{\#PizzaHut \#PizzaHutMalaysia \#HolidayDeals\\ 
\#HutLovers \#MyBoxEverydayContest} \\
\hline
Burger King & \makecell[l]{\#VoteForChange \#burgerking \#BurgerKingMY\\ 
\#LuckyYourWay \#BurgerElection} \\
\bottomrule
\end{tabular}
\end{table}

The hashtags used by outlets in developing countries, shown in \autoref{tab:hashtags_developing}, indicate a strong and consistent focus on localization and cultural integration. For instance, Starbucks uses regional tags like \#StarbucksMalaysia and Spanish phrases like \#ViveTuAhora. Subway directly incorporates local language with \#MakanFreshSubway (Malaysia), while KFC targets specific nations with \#KFCIndonesia and \#KFCArabia. This pattern is consistent across nearly all brands, including McDonald's with \#McDonaldsTürkiye and Pizza Hut with \#PizzaHutMalaysia. Overall, the dominant strategy in these regions is to establish a strong local presence and connect with national identity, a clear contrast to the more niche-focused strategies seen in developed markets.

\subsubsection{\textbf{Image Analysis}}

In this phase of our study, we analyzed the visual content from Instagram posts of major food brands, focusing specifically on extracting the dominant colors within the images. This analysis was conducted separately for developed and developing countries to uncover patterns in the color schemes employed by these brands across different economic contexts.

To extract the dominant colors, we used KMeans clustering, a machine learning algorithm that divides data points into clusters. Each image was reduced in resolution and converted into RGB format. The pixel data was flattened into a two-dimensional array and KMeans clustering was applied to group the pixels into five clusters, each representing a dominant color. This method effectively groups similar colors together by minimizing the difference within each cluster and identifying the centroids of the clusters as the representative colors. The formula for KMeans clustering minimizes the sum of squared distances between each point (pixel) and its nearest centroid (dominant color):

\[
\arg \min_{S} \sum_{i=1}^{k} \sum_{x \in S_i} || x - \mu_i ||^2
\]

Here, \( k \) represents the number of colors (clusters) extracted from the image, \( S_i \) is the set of pixels assigned to the \( i^{th} \) color, and \( \mu_i \) is the centroid, or the representative color, of that cluster.

Once the dominant colors were identified, they were displayed in bands, with the size of each color band representing the relative dominance of that color in the image. In other words, the larger the area occupied by a color in the band, the more dominant it was for that outlet. By aggregating the color data across all posts for each outlet, we established the most commonly used color schemes in both developed and developing regions. This approach allowed for a consistent comparison of visual strategies employed by brands across different geographical markets.

\begin{figure}[htp]
\centering
\begin{tabular}{|c|c|}
\hline
\textbf{Developed} & \textbf{Developing} \\
\hline
\includegraphics[width=0.45\textwidth]{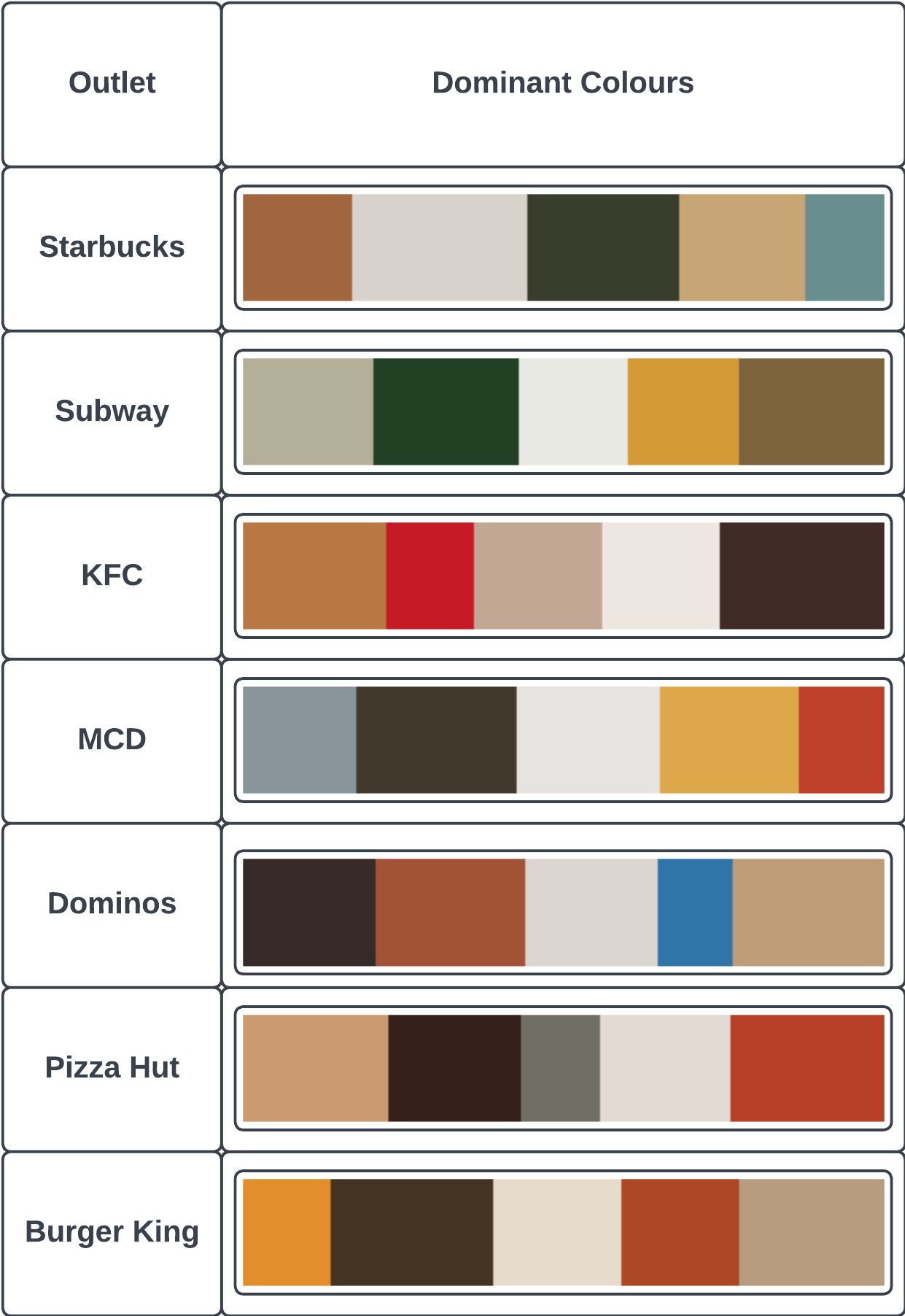} &
\includegraphics[width=0.45\textwidth]{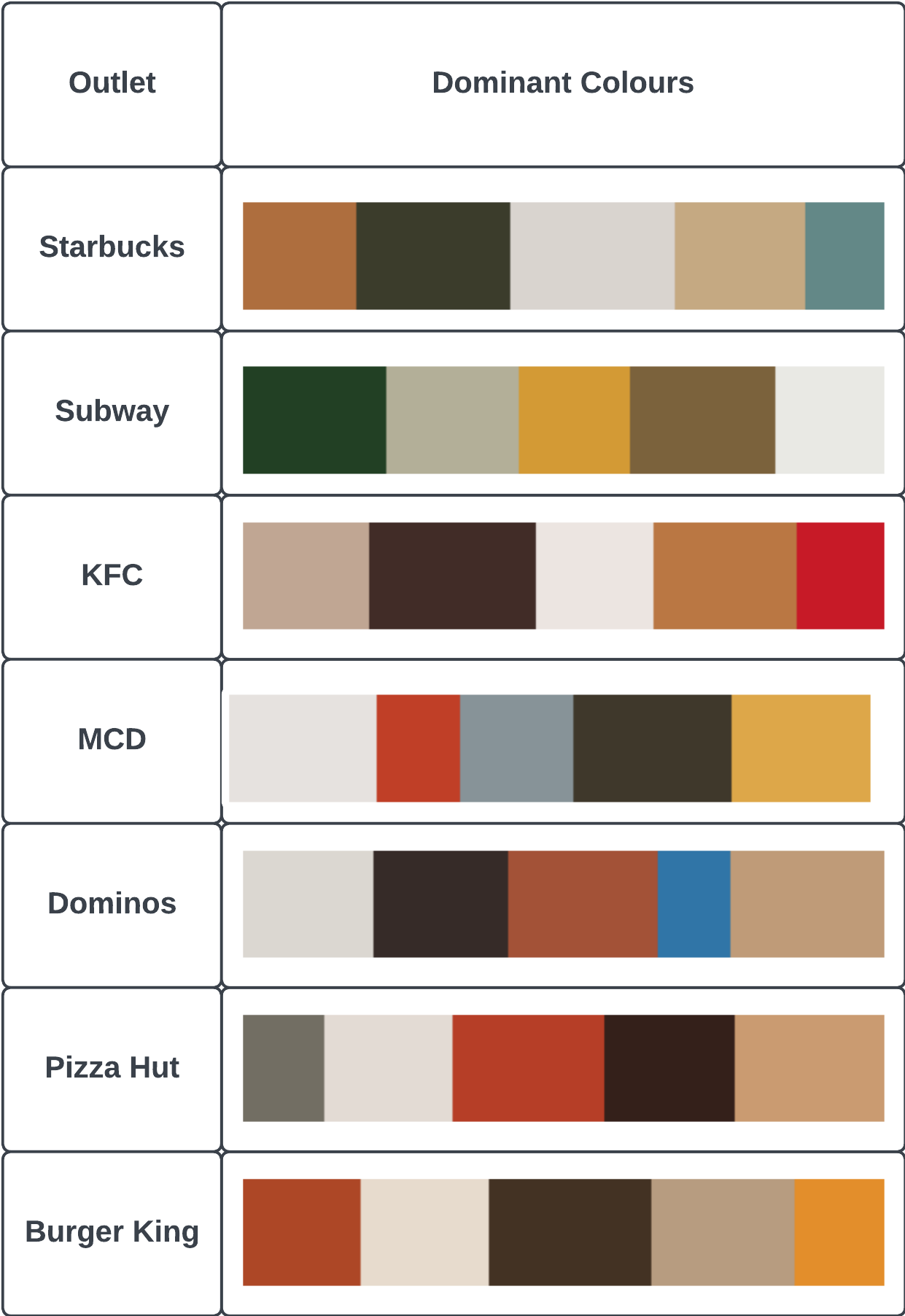} \\
\hline
\end{tabular}
\caption{Dominant Colours extracted from developed and developing countries}
\label{fig:dominant_colours_comparison}
\end{figure}

\textbf{1. Developed Countries : } \\
For the developed countries, we extracted the five most dominant colors for each outlet's Instagram posts to gain insights into their visual marketing strategies. As shown in \autoref{fig:dominant_colours_comparison}, these dominant colors represent the primary hues that brands utilize in their images, with a focus on visual consistency and brand recognition.

For instance, Starbucks employs a mix of earthy tones, including browns, greens, and beige, which resonate with the brand's natural and sophisticated image. Subway similarly uses green and beige tones, aligning with its fresh and healthy brand identity. KFC incorporates a bold red in its palette, indicative of its iconic branding. MCD's dominant colors include shades of yellow and red, synonymous with its global brand identity. Dominos, Pizza Hut, and Burger King feature darker, neutral shades, with occasional brighter colors like blue and orange to attract attention.

\textbf{2. Developing Countries : } \\
For the developing countries, we similarly extracted the five most dominant colors for each outlet's Instagram posts to identify visual marketing strategies tailored for these markets. \autoref{fig:dominant_colours_comparison} illustrates the dominant colors for popular food brands in developing regions, offering insights into how these brands visually communicate with their audience.

For Starbucks, the colors include earthy and muted tones, such as browns and greens, reflecting its consistent global brand image. Subway emphasizes the use of natural green alongside yellow and beige, aligning with its fresh, healthy message. KFC's palette prominently features red, echoing its bold and distinctive branding, while MCD continues to use its signature yellow and red, indicative of its universally recognizable color scheme. Dominos, Pizza Hut, and Burger King favor darker hues with occasional splashes of bright colors like red or blue to stand out.

\subsection{\textbf{Regression Analysis of Structural Factors}}
To complement the multimodal analysis, we conducted a regression-based investigation to assess how macro-level geographic and health factors—specifically population size,GDP , and obesity prevalence—influence Instagram engagement metrics across developed and developing countries.

\subsubsection{Data Collection and Variables}

Structural data on Population, GDP(in current USD), and Obesity Rate (percentage of the adult population classified as obese) were obtained from the World Population Review (2024). These indicators were matched with Instagram post data from national accounts of five global food brands—MCD, KFC, Domino’s,Subway and Starbucks—across multiple countries.

Dependent variables included three engagement metrics at the post level: \textit{Likes}, \textit{Comments}, \textit{Views}. Independent variables—\textbf{Population} and \textbf{GDP}—were log-transformed to correct for skewness. 
\subsubsection{Model Design}

We estimated separate multiple linear regression models for each engagement metric, stratified by development status (developed vs. developing). The model specification was:

\[
\text{Engagement Metric}_i = \beta_0 + \beta_1 \text{Population}_i + \beta_2 \text{GDP}_i + \beta_3 \text{Obesity}_i + \epsilon_i
\]

Where \( i \) indexes individual posts. All models were estimated using Ordinary Least Squares (OLS). Segmenting by region allowed us to examine how structural factors influence digital engagement in different economic contexts.

\section{Results}
In this section, we present the findings from our analysis, focusing on the relationships between engagement metrics and various factors such as geographic context, dominant colors, and sentiment. Using a series of ANOVA tests, we examined whether significant differences exist across these dimensions, building upon the findings from our methodology to explore correlations with consumer engagement.

\subsection{\textbf{Engagement of Outlets in developed vs developing countries}} 

We first analyzed whether overall engagement differs significantly for each brand between developed and developing countries. Using a one-way ANOVA test for each outlet, we compared the mean engagement metrics (likes, comments, and views) across the two regional categories to determine if the economic context impacts consumer interaction.

\begin{table}[h]
\centering
\caption{Comparison of engagement metrics between developed and developing countries for each outlet}
\renewcommand{\arraystretch}{1.2} 
\setlength{\tabcolsep}{9pt}      

\label{corr1}
\begin{tabular}{@{}ll@{}}
\toprule
\textbf{Outlet} & \textbf{p-value} \\
\midrule
Starbucks & 0.026022 \\
\hline
MCD & 0.503845 \\
\hline
KFC & 0.271562 \\
\hline
Pizza Hut & 0.048596 \\
\hline
Burger King & 0.048082 \\
\hline
Domino's & 0.485994 \\
\hline
Subway & 0.876218 \\
\bottomrule
\end{tabular}
\end{table}

The results, presented in \autoref{corr1}, show the p-values from this comparison. Significant differences in engagement were found for Starbucks (p=0.026), Pizza Hut (p=0.048), and Burger King (p=0.048), indicating that these brands' content resonates differently depending on the economic context. Conversely, MCD, KFC, Domino's, and Subway did not show statistically significant differences in their overall engagement between the two regions. This suggests that the audience interaction with these latter brands is more consistent across the developed and developing markets included in our study.

\subsection{\textbf{Dominant Colours vs Engagement}}

To investigate the influence of visual aesthetics, we next explored whether the dominant colors in post images significantly impact engagement. ANOVA tests were performed separately for developed and developing countries, revealing a striking difference between the two contexts.

In developed countries, we found no evidence that dominant colors affect engagement. As shown in \autoref{T1}, the p-values for all engagement metrics (Likes, Comments, Followers, Views, and Total Interactions) were well above the 0.05 significance threshold. This null result suggests that for audiences in these mature markets, the specific color palette of a post may be less influential on their interaction behavior.

\begin{table}[h]
\caption{ANOVA Results for Engagement Metrics by Dominant Colour (Developed Countries)}
\renewcommand{\arraystretch}{1.2} 
\setlength{\tabcolsep}{9pt}      
\small\sf\centering
\begin{tabular}{lcc}
\toprule
\textbf{Metric} & \makecell{\textbf{F-} \\ \textbf{statistic}} & \textbf{p-value} \\
\midrule
Likes & 0.186725 & 0.926067 \\
\hline
Comments & 0.487944 & 0.756064 \\
\hline
Followers & 1.069687 & 0.535605 \\
\hline
Views & 1.143321 & 0.515949 \\
\hline
Total Interactions & 0.188414 & 0.925092 \\
\bottomrule
\end{tabular}
\label{T1}
\end{table}

In stark contrast, dominant colors were found to be a significant factor in developing countries. The ANOVA results in \autoref{T2} show a statistically significant relationship between color and Likes (p=0.043), with a near-significant trend for Total Interactions (p=0.051). This indicates that in these regions, visual color schemes play a more critical role in attracting user engagement.

\begin{table}[h]
\caption{ANOVA Results for Engagement Metrics by Dominant Colour (Developing Countries)}
\small\sf\centering
\renewcommand{\arraystretch}{1.2} 
\setlength{\tabcolsep}{9pt}      
\begin{tabular}{lcc}
\toprule
\textbf{Metric} & {\textbf{F-statistic}} & \textbf{p-value} \\
\midrule
Likes & 22.239234 & 0.043493 \\
\hline
Comments & 1.352703 & 0.466919 \\
\hline
Followers & 0.946081 & 0.571973 \\
\hline
Views & 8.358526 & 0.109700 \\
\hline
Total Interactions & 18.633150 & 0.051582 \\
\bottomrule
\end{tabular}
\label{T2}
\end{table}

To pinpoint exactly which colors were driving this effect on Likes, we conducted a post-hoc Tukey's HSD (Honestly Significant Difference) test. The results, summarized in \autoref{tab:Tukeys}, reveal that posts featuring a combination of "Off-white" and "Green" received a significantly different number of Likes compared to other color combinations (p=0.0497). No other specific color pairings showed a statistically significant effect.

\begin{table}[h]
\caption{Tukey HSD Test for Colour Pairs Affecting Likes (Developing Countries)}
\small\sf\centering
\begin{tabular}{lcc}
\toprule
\textbf{Colour Pair} & \textbf{p-value} & \textbf{Sig Difference} \\
\midrule
Off white vs Green & 0.0497 & True \\
\hline
Light brown vs Off white & 0.0541 & False \\
\hline
Light brown vs Brown & 0.1215 & False \\
\hline
Light brown vs Green & 0.9953 & False \\
\hline
Light brown vs Red & 0.7055 & False \\
\hline
Off white vs Brown & 0.1502 & False \\
\hline
Off white vs Red & 0.0791 & False \\
\hline
Brown vs Green & 0.1048 & False \\
\hline
Brown vs Red & 0.2498 & False \\
\hline
Green vs Red & 0.5806 & False \\
\bottomrule
\end{tabular}
\label{tab:Tukeys}
\end{table}

These findings collectively suggest that while color strategy may have a negligible impact in developed markets, it is a key lever for driving engagement in developing countries. Specifically, the successful combination of off-white and green tones points towards a visual aesthetic that resonates strongly with audiences in these emerging digital landscapes.

\subsection{\textbf{Sentiments vs Engagement}} 
We examined the relationship between the sentiment of post descriptions (Positive, Neutral, or Negative) and user engagement. The analysis, conducted separately for developed and developing countries, reveals that sentiment has a complex but significant influence, most notably on the size of a brand's audience.

Our most consistent finding is a strong, near-universal correlation between post sentiment and follower count. As shown in \autoref{tab:T3} and \autoref{tab:T4}, the sentiment of a brand's posts is a highly significant predictor of its follower base (p<0.01) for almost every brand across both developed and developing regions. This suggests that a consistent emotional tone in messaging is a critical factor in building and maintaining a large audience on the platform. The only exception was MCD in developed countries, where no such significant relationship was found.

\begin{table*}[h]
\caption{ANOVA Results for Engagement Metrics by Sentiment (Developed Countries)}\label{tab:T3}
\centering
\small\sf
\resizebox{1\textwidth}{!}{
\begin{tabular}{lccccc}
\toprule
\textbf{Outlet} & \textbf{Likes} & \textbf{Comments} & \textbf{Followers} & \textbf{Views} & \textbf{Total Interactions} \\
\midrule
Starbucks & 0.06 & 0.68 & **3.11e-06 & 0.67 & 0.05*\textsuperscript{a} \\ 
\hline
Subway & 0.08 & 0.59 & **4.38e-05 & 0.55 & 0.08 \\
\hline
KFC & **4.82e-05 & 0.54 & **5.6e-07 & 0.63 & **4.7e-05 \\
\hline
MCD & *0.007 & 0.99 & 1.83 & 0.7 & *0.006 \\
\hline
Domino's & 0.2 & 0.95 & **1.37e-05 & 0.56 & 0.16 \\
\hline
Pizza Hut & 0.35 & 0.64 & **3.2e-04 & 0.61 & 0.33 \\
\hline
Burger King & *0.02 & 0.56 & **5.5e-06 & 0.62 & *0.02 \\
\bottomrule
\end{tabular}}
\vspace{0.5cm} 
\noindent\footnotesize{\textsuperscript{a} Significance levels: * $p < 0.05$, ** $p < 0.01$.}
\end{table*}

In contrast, the impact of sentiment on immediate, post-level engagement metrics like Likes and Total Interactions is more varied. In developed countries (\autoref{tab:T3}), brands like KFC (p<0.01), MCD (p<0.01), and Burger King (p<0.05) saw a significant relationship between sentiment and both Likes and Total Interactions. However, this effect was not significant for other major brands like Starbucks or Domino's.

A similar, though less pronounced, pattern was observed in developing countries (\autoref{tab:T4}). Here, only KFC (p<0.05) showed a significant link between sentiment and immediate engagement (Likes and Total Interactions), while this relationship was not significant for the other brands in this region.

\begin{table*}[h]
\caption{ANOVA Results for Engagement Metrics by Sentiment (Developing Countries)}
\centering
\small\sf
\resizebox{1\textwidth}{!}{
\begin{tabular}{lccccc}
\toprule
\textbf{Outlet} & \textbf{Likes} & \textbf{Comments} & \textbf{Followers} & \textbf{Views} & \textbf{Total Interactions} \\
\midrule
Starbucks & 0.07 & 0.69 & **1.07e-06 & 0.99 & 0.06 \\
\hline
Subway & 0.82 & 0.62 & **4.9e-05 & 0.61 & 0.80 \\
\hline
KFC & *0.04 & 0.7 & **1.75e-07 & 0.77 & *0.04 \\
\hline
MCD & 0.16 & 0.68 & **5.4e-07 & 0.76 & 0.14 \\
\hline
Domino's & 0.65 & 0.72 & **2.39e-06 & 0.59 & 0.51 \\
\hline
Pizza Hut & 0.38 & 0.69 & **1.0e-06 & 0.97 & 0.33 \\
\hline
Burger King & 0.16 & 0.65 & **3.6e-06 & 0.78 & 0.15 \\
\bottomrule
\end{tabular}}
\vspace{0.5cm} 
\noindent\footnotesize{\textsuperscript{a} Significance levels: * $p < 0.05$, ** $p < 0.01$.}
\label{tab:T4}
\end{table*}

Overall, these results indicate that while the emotional tone of a single post does not guarantee immediate likes or interactions for every brand, a strategically managed sentiment profile is a powerful factor associated with building a large and loyal follower base across diverse economic contexts.

\subsection{\textbf{Outlets vs Engagement}}

Before examining our regression models, we sought to confirm whether brand identity itself is a primary driver of engagement, regardless of geographic context. To do this, we conducted a one-way ANOVA test, using the pooled data from all regions to compare the mean engagement metrics across the seven food outlets.

The results, presented in \autoref{tab:T5}, are unequivocal. We found highly significant differences across all engagement metrics, including Likes (p<0.001), Comments (p<0.001), Views (p=0.005), Followers (p<0.001), and Total Interactions (p<0.001). This statistically confirms the significant performance variations observed in our earlier descriptive analysis—where brands like KFC, Starbucks, and MCD demonstrated distinct leadership in different contexts. This foundational finding establishes that the specific brand is a crucial factor in determining engagement levels, reinforcing the importance of analyzing the unique strategies that drive their varied success.

\begin{table}[h]
\caption{ANOVA Results for Engagement Metrics by Outlets}
\small\sf\centering
\resizebox{0.5\textwidth}{!}{
\begin{tabular}{lc}
\toprule
\textbf{Metric} & \textbf{p-value} \\
\midrule
Likes avg & 0.00000 \\
\hline 
Comments avg & 0.00000 \\
\hline 
Views avg & 0.00540 \\
\hline 
Followers avg & 0.00000 \\
\hline 
Total Interactions avg & 0.00000 \\
\bottomrule
\end{tabular}}
\label{tab:T5}
\end{table}

\subsection{\textbf{Regression Results: Structural Influences on Engagement}}

Finally, our regression analysis reveals how country-level structural factors shape engagement, often in directly opposing ways depending on the economic context. The results for both developed and developing countries are detailed in \autoref{tab:regression_engagement} and \autoref{tab:regression_engagement_alt}, respectively.

Population: The effect of a country's population shows a stark reversal between regions. In developed countries, a larger population is a significant positive predictor of both Likes (Coef = 816.99, p<0.001) and Views (Coef = 104.03, p<0.001), suggesting that a larger audience pool directly translates to higher engagement. Conversely, in developing countries, a larger population is associated with a significant decrease in both Likes (Coef = -2.28, p<0.001) and Comments (Coef = -0.23, p<0.001).

GDP: National wealth also exhibits opposing effects. In developed countries, a higher GDP has a significant negative impact on engagement, reducing both Likes (Coef = -2.07e+04, p<0.001) and Views (Coef = -1746.85, p=0.01). This aligns with theories of market saturation. In striking contrast, for developing countries, a higher GDP is a powerful positive predictor of engagement, significantly increasing both Likes (Coef = 2012.45, p<0.001) and Comments (Coef = 73.51, p<0.001).

\begin{table}[h]
\caption{Regression Results: Impact of Population, GDP, and Obesity Rate on Engagement Metrics (Developed Countries)}
\small\sf\centering
\begin{tabular}{lcccccc}
\toprule
\textbf{Metric} & \multicolumn{2}{c}{\textbf{Likes}} & \multicolumn{2}{c}{\textbf{Comments}} & \multicolumn{2}{c}{\textbf{Views}} \\
\cmidrule(lr){2-3} \cmidrule(lr){4-5} \cmidrule(lr){6-7}
\textbf{Variable} & \textbf{Coef} & \textbf{p-value} & \textbf{Coef} & \textbf{p-value} & \textbf{Coef} & \textbf{p-value} \\
\midrule
Population (millions) & 816.992 & 0.0000 & 1.0295 & 0.598 & 104.03 & 0.000 \\
GDP (trillions) & -2.07e+04 & 0.000 & -50.54 & 0.327 & -1746.85 & 0.010 \\
Obesity Rate (\% of population) & 258.85 & 0.009 & -5.567 & 0.070 & 37.6518 & 0.355 \\
\bottomrule
\end{tabular}
\label{tab:regression_engagement}
\end{table}

Obesity Rate: The influence of national obesity rates is more nuanced. In developed countries, a higher obesity rate is positively associated with a significant increase in Likes (Coef = 258.85, p=0.009). In developing countries, however, the obesity rate has no significant effect on Likes but is linked to a significant decrease in Comments (Coef = -2.99, p<0.001).

These contrasting results strongly underscore that the structural drivers of social media engagement are not universal. Factors like population and economic wealth function in fundamentally different ways depending on the maturity and dynamics of the market.

\begin{table}[h]
\caption{Regression Results: Effect of Population, GDP, and Obesity Rate on Engagement Metrics (Developing Countries)}
\small\sf\centering
\begin{tabular}{lcccccc}
\toprule
\textbf{Metric} & \multicolumn{2}{c}{\textbf{Likes}} & \multicolumn{2}{c}{\textbf{Comments}} & \multicolumn{2}{c}{\textbf{Views}} \\
\cmidrule(lr){2-3} \cmidrule(lr){4-5} \cmidrule(lr){6-7}
\textbf{Variable} & \textbf{Coef} & \textbf{p-value} & \textbf{Coef} & \textbf{p-value} & \textbf{Coef} & \textbf{p-value} \\
\midrule
Population (millions) & -2.2841 & 0.000 & -0.229 & 0.000 & -0.042 & 0.870 \\
GDP (trillions) & 2012.448 & 0.000 & 73.51 & 0.000 & 58.58 & 0.587 \\
Obesity Rate (\% of population) & -4.0620 & 0.559 & -2.99 & 0.000 & -3.77 & 0.434 \\
\bottomrule
\end{tabular}
\label{tab:regression_engagement_alt}
\end{table}

\section{Discussion}
This study examined multimodal Instagram marketing strategies, revealing that geographic context, dominant colors, and sentiment significantly impacted consumer engagement. Brands like Starbucks and Pizza Hut experienced varied engagement depending on regional context, while colors such as off-white and green boosted interaction in developing countries. Additionally, sentiment played a crucial role, with MCD and Starbucks receiving more positive feedback, whereas KFC and Domino's saw higher negative sentiment. Conversely, globally consistent strategies, as seen with MCD and KFC, maintained steady performance across regions. Additionally, our regression analysis revealed that structural factors such as GDP and population significantly shape engagement outcomes, with contrasting patterns across developed and developing countries.These findings underscore the importance of adapting marketing strategies to cultural and economic factors, with locally tailored approaches often yielding stronger engagement. 

The differences in engagement between developed and developing countries for brands such as Starbucks and Pizza Hut can be attributed to their localized marketing strategies. MCD has successfully adapted to local cultures by offering region-specific menu items—such as the vegetarian McAloo Tikki in India, Teriyaki McBurger in Japan, and Croque McDo in France—to reflect local tastes and dietary preferences \citep{mcdomenuguideMcDonaldsAdapts}. Similarly, Pizza Hut customizes its menu to local preferences, offering vegetarian options such as paneer pizza in India and avoiding pork in Muslim-majority regions. The brand also adjusts its pricing strategies to emphasize value deals in regions with lower disposable incomes, ensuring broader accessibility \citep{thestrategystoryPizzaPESTEL}.
The influence of color on consumer engagement is well-documented, with colors playing a critical role in shaping perceptions and driving purchasing decisions \citep{bytycci2020influence}. Studies have shown that brighter, more saturated images boost post popularity on platforms like Instagram \citep{yu2020coloring}, while darsker tones like black and gray increase engagement on Facebook \citep{zailskaite2017brand}. In developing regions, where brand loyalty is less established, visually striking content may play a more pivotal role in capturing consumer attention \citep{han2023perceived}. Vibrant colors such as red and yellow, known to drive emotional responses, especially among younger consumers like Gen Z, are particularly effective in boosting engagement \citep{zaiddi2023effectiveness}.Studies in marketing psychology show that while the color green is widely used to convey nature, health, and freshness, off-white tones are employed to evoke a sense of simplicity, cleanliness, and modern sophistication \citep{aslam2006you,labrecque2012exciting}. Our findings support this, with dominant colors like off-white and green correlating with higher Likes and Total Interactions, underscoring the importance of visual appeal in these regions where trust is still being built \citep{ren2018influence}.
Sentiment and tone play a crucial role in influencing consumer engagement on social media. Emotionally charged content tends to generate more interactions, as consumers connect with posts that evoke emotions like joy, excitement, nostalgia, driving key engagement metrics like Likes, Shares, and Follower growth \citep{singlegrainAppealEmotions, kim2019emotional}. Emotional content creates lasting impressions, strengthening brand loyalty and enhancing digital marketing success \citep{noblestudiosPowerEmotional, freedomtoascendRoleEmotional}. 
Interestingly, negative emotional appeals, particularly in charity and environmental campaigns, also drive engagement by urging consumers to take action \citep{yousef2021social}. Our findings reflect this, showing a particularly strong and consistent correlation between a brand's post sentiment and its follower growth. The impact on immediate post-level metrics like Likes and Interactions was significant for some brands but not universal, suggesting that sentiment's primary role may be in long-term audience building.

Beyond visual and emotional elements, our study highlights the significant role of structural factors—specifically population size, GDP and obesity rates—in shaping Instagram engagement across different regional contexts. Prior research suggests that higher national populations generally increase social media engagement by expanding the pool of potential users and participants in online interactions \citep{datareportalGlobalSocial}. Our findings in developed countries align with this, showing that population positively influences likes and views. Our regression analysis reveals that a country's GDP plays a dual and contrasting role in shaping social media engagement, a trend that aligns with existing research. In developed countries, we observed a negative correlation between GDP and engagement metrics. This can be attributed to market saturation and high levels of ad exposure in wealthier economies, which can lead to consumer disengagement and ad fatigue \citep{koltan2013market,fernandes2024brands}.Conversely, in developing regions, GDP positively predicts engagement metrics like likes and comments. This dynamic is driven by rising disposable income and increased digital access, which enable enthusiastic consumers to interact with aspirational global brands. This effect is often amplified by the novelty of their newfound ability to participate in the global marketplace, leading to higher engagement rates \citep{sheth2011impact,steenkamp2010global,BATRA200083}.Regarding health indicators, previous work links obesity prevalence to patterns of food-related social media engagement, emphasizing that cultural factors and food marketing strategies interact strongly with consumer responsiveness \citep{aljefree2022exposure, gu2021associations,sadek2024digital,tazeouglu2022effect}. While existing research does not differentiate the obesity-engagement relationship by economic context, our study adds new insight by showing how obesity rates exhibit different engagement effects across regions.

The findings of this study have significant implications for global marketing strategies. Brands operating across diverse geographic regions must recognize that a one-size-fits-all approach is insufficient. Instead, successful marketing requires a nuanced understanding of regional differences in cultural values, economic conditions, and consumer behaviors. By integrating multimodal elements—visual aesthetics, emotional tone, and structural factors—into their marketing strategies, brands can create more personalized and effective campaigns. This tailored approach not only enhances consumer engagement but also fosters stronger brand loyalty and long-term success in the global marketplace.

While this study provides valuable insights into multimodal marketing strategies, it has certain limitations. The analysis focused solely on Instagram posts, excluding other formats like Instagram Reels or Stories, which are increasingly popular. Additionally, the study did not account for other visual elements such as image composition, typography, or the presence of logos, which could further influence consumer engagement. Future research should consider a broader range of social media platforms and content types, as well as a more comprehensive set of visual and textual elements. Moreover, longitudinal studies examining the long-term effects of tailored marketing strategies on brand loyalty and consumer behavior would provide deeper insights into the efficacy of these approaches.

\section{Conclusion}

This study delivers a novel, dual-lens analysis of Instagram marketing, demonstrating that user engagement is shaped not only by multimodal content features but also by structural, country-level factors. Through a combination of sentiment analysis, hashtag extraction, and dominant color detection, we revealed how visual aesthetics and emotional tone drive engagement differently across global food brands. For instance, we found that specific color palettes like off-white and green significantly boost interactions in developing countries, a context where such visual cues are surprisingly potent. Furthermore, our regression analysis confirmed that macroeconomic indicators have powerful, often opposing, effects on engagement: a higher GDP, for example, correlates with reduced attention in developed nations but strongly predicts positive engagement in developing ones.

By bridging content-level insights with macroeconomic and health-related dimensions, this research extends the literature on social media engagement and offers actionable implications for global brands. The core takeaway is the critical need to move beyond a one-size-fits-all approach and develop localized digital marketing strategies that align both content design and contextual realities.

Future research should build on the limitations of this framework, for example by incorporating increasingly popular modalities like Instagram Reels and Stories, or by analyzing a more comprehensive set of visual elements like image composition and typography. Expanding the temporal and platform scope of this work will also be crucial for capturing the evolving dynamics of engagement in an increasingly globalized digital landscape. Ultimately, understanding this interplay between design and structure is essential for marketers seeking to craft more effective and culturally responsive campaigns.

\appendix
\section*{Appendix A}

This appendix provides a detailed breakdown of the Instagram post data attributes collected via the CrowdTangle API, as referenced in ~\autoref{sec:data_collection}.

\begin{table}[H]
\caption{Instagram Post Data Attributes\label{tab:instagram_post_data_attributes}}

\newlength{\extralength}
\setlength{\extralength}{1cm}

\begin{adjustwidth}{-\extralength}{0cm}

\begin{tabularx}{\linewidth}{l l X} 
\toprule
\textbf{Attribute} & \textbf{Datatype} & \textbf{Attribute Description} \\
\midrule
Account & String & The Instagram handle of the food outlet. \\ \hline 
User Name & String & The name of the user or the Instagram handle. \\ \hline 
Followers at Posting & int & The number of followers at the time of posting. \\ \hline 
Post Created & date-time & The timestamp of when the post was created. \\ \hline 
Post Created Date & date-time & The date of post creation. \\ \hline 
Post Created Time & date-time & The time of post creation. \\ \hline 
Type & String & The type of post (e.g., image, video, carousel). \\ \hline 
Total Interactions & int & The sum of likes and comments the post received.  \\ \hline 
Likes & int & The number of likes the post received. \\ \hline 
Comments & int & The number of comments on the post. \\ \hline 
Views & int & The number of views for video posts. \\ \hline 
Like and View Counts Disabled & boolean & Indicator if like and view counts were disabled. \\ \hline 
URL & String & The URL of the post. \\ \hline 
Link & String & Any external link included in the post. \\ \hline 
Photo & String & URL of the photo in the post. \\ \hline 
Title & String & The title of the post. \\ \hline 
Description & String & The caption or description of the post. \\ \hline 
Image Text & String & Any text detected in the image. \\ \hline 
Sponsor Id & String & The ID of the sponsor, if any. \\ \hline 
Sponsor Name & String & The name of the sponsor. \\ \hline 
Overperforming Score & int & A score indicating how well a post's total interactions performed relative to the account's typical total interactions for that type of post. \\
\bottomrule
\end{tabularx}

\end{adjustwidth}

\end{table}

\section*{Appendix B}
The following graphs illustrate country-wise distributions of engagement metrics—likes, comments, views, followers, total interactions, and overperforming score—alongside structural factors such as population and GDP. Each plot is grouped by outlet and categorized into developed and developing regions. The y-axis is presented on a logarithmic scale to facilitate visual comparison across metrics with large magnitude differences. These plots provide a comparative snapshot of how engagement levels and structural indicators vary across national contexts.

\setlength{\extralength}{0cm} 

\begin{table}[H]
    \begin{adjustwidth}{-\extralength}{0cm} 
        \begin{tabularx}{\linewidth}{m{1.5cm} X}
            \toprule
            \textbf{Outlets} & \textbf{Developed Countries} \\
            \midrule
            
            \scriptsize MCD &
            \includegraphics[width=\linewidth]{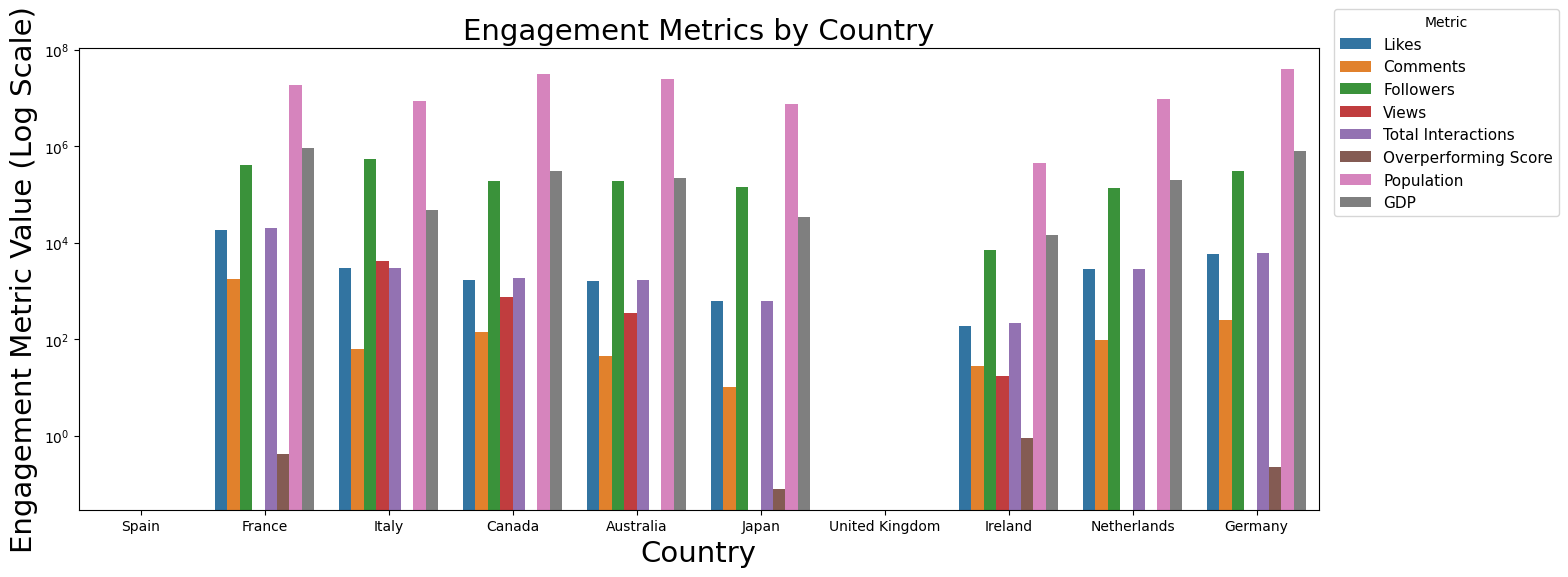} \\
            
            \addlinespace[2em]
            
            \scriptsize Starbucks &
            \includegraphics[width=\linewidth]{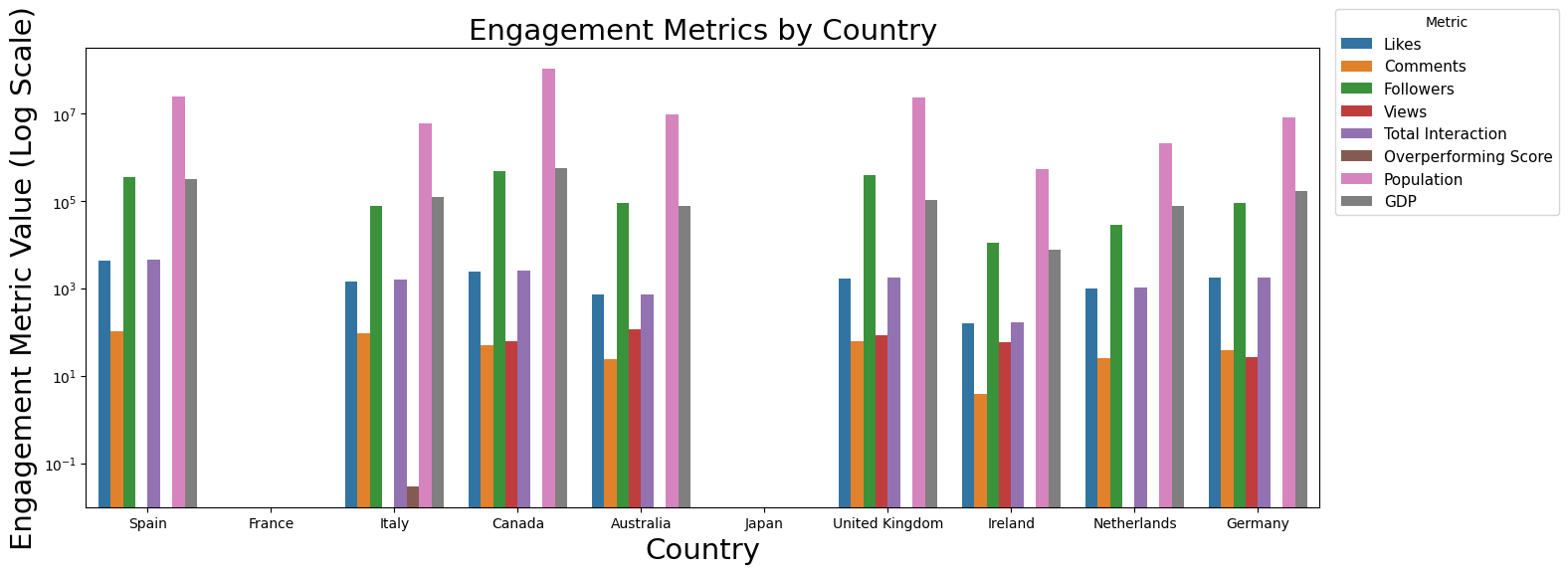} \\

            \addlinespace[2em]
            
            \scriptsize BurgerKing &
            \includegraphics[width=\linewidth]{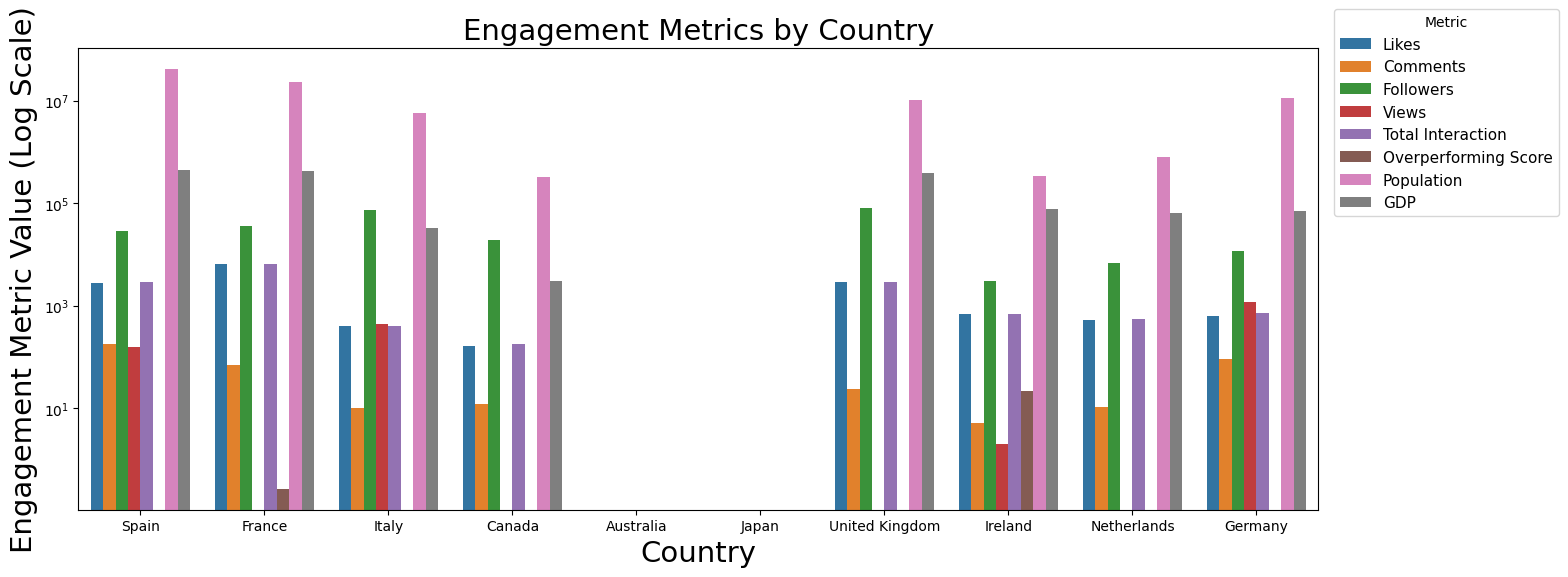} \\
            
            \bottomrule
        \end{tabularx}
    \end{adjustwidth} 
\end{table}

\clearpage

\begin{table}[H]
    \begin{adjustwidth}{-\extralength}{0cm} 
        \begin{tabularx}{\linewidth}{m{1.5cm} X}
            \toprule
            \textbf{Outlets} & \textbf{Developed Countries} \\
            \midrule
            
            \scriptsize Dominos &
            \includegraphics[width=\linewidth]{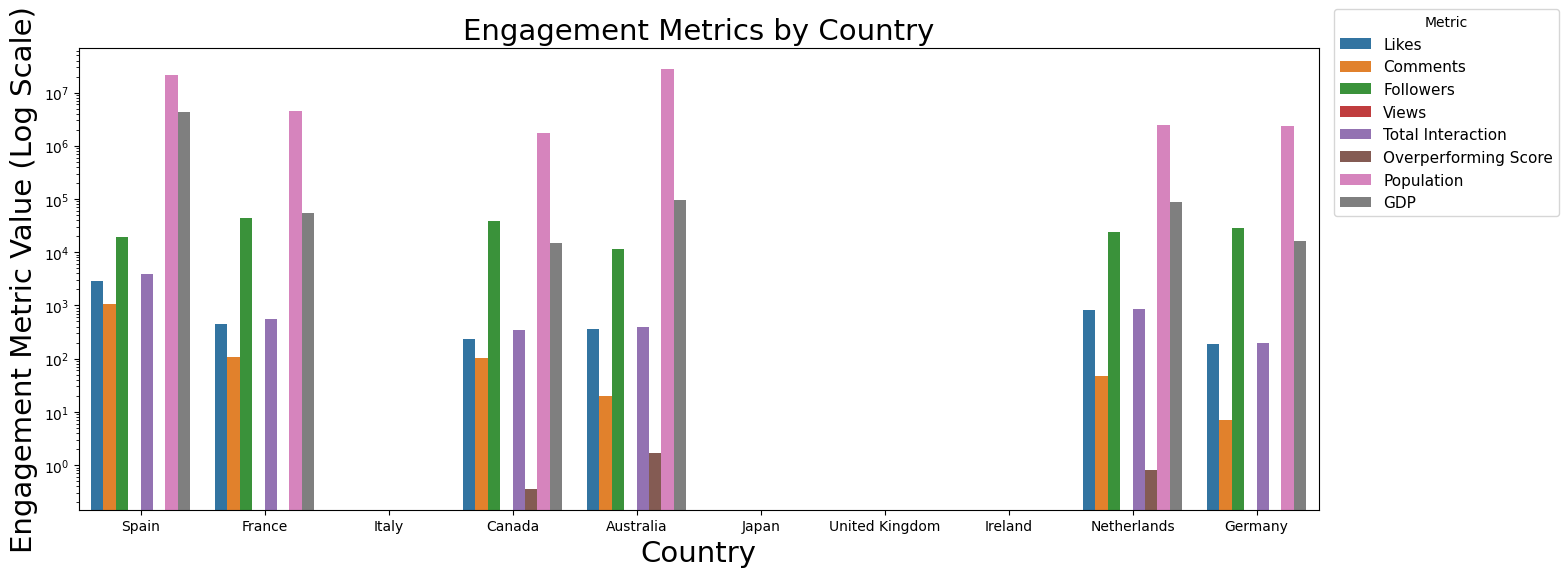} \\
            
            \addlinespace[2em]
            
            \scriptsize PizzaHut &
            \includegraphics[width=\linewidth]{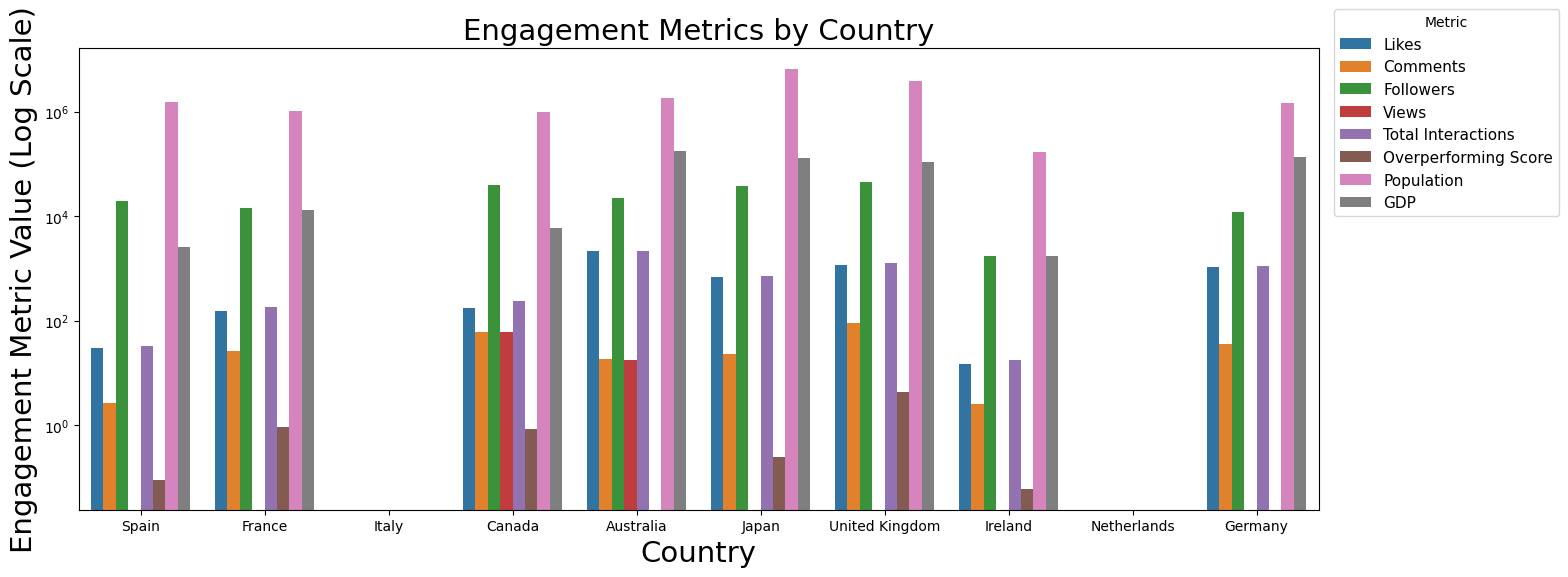} \\

            \addlinespace[2em]
            
            \scriptsize KFC &
            \includegraphics[width=\linewidth]{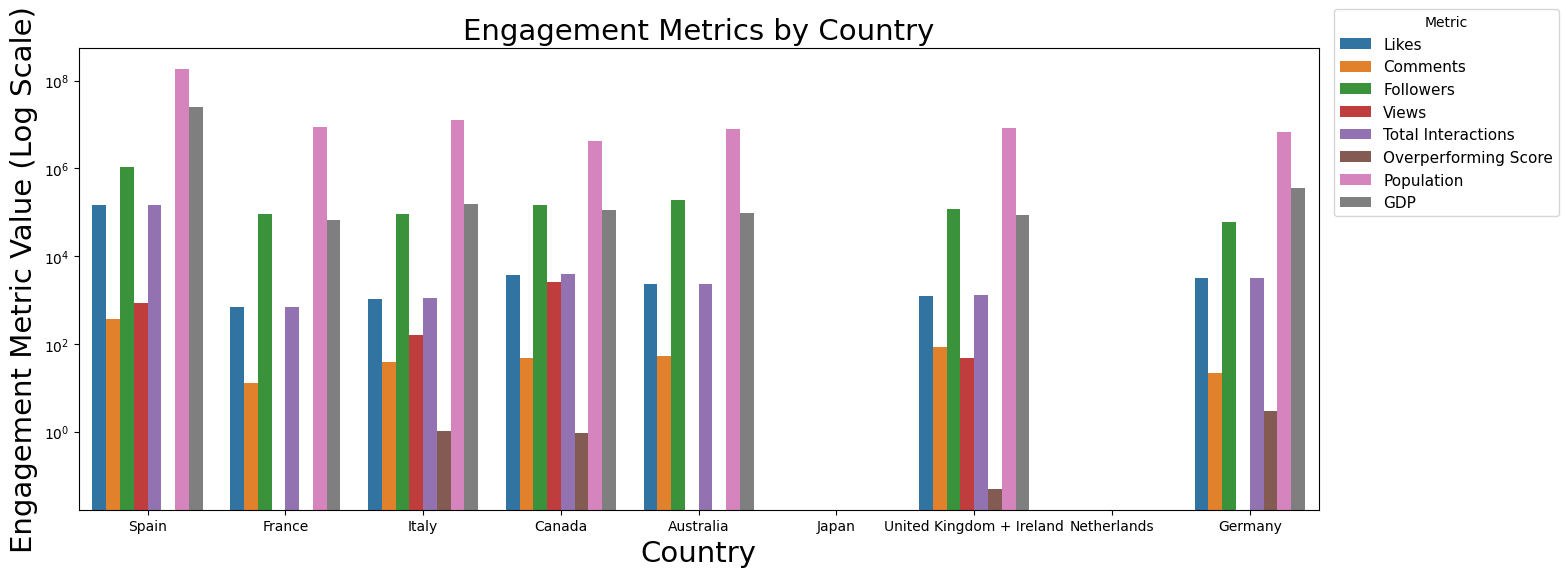} \\
            
            \bottomrule
        \end{tabularx}
    \end{adjustwidth}
\end{table}

\clearpage

\begin{table}[H]
    \begin{adjustwidth}{-\extralength}{0cm} 
        \begin{tabularx}{\linewidth}{m{1.5cm} X}
            \toprule
            \textbf{Outlets} & \textbf{Developed Countries} \\
            \midrule
            
            \scriptsize Subway &
            \includegraphics[width=\linewidth]{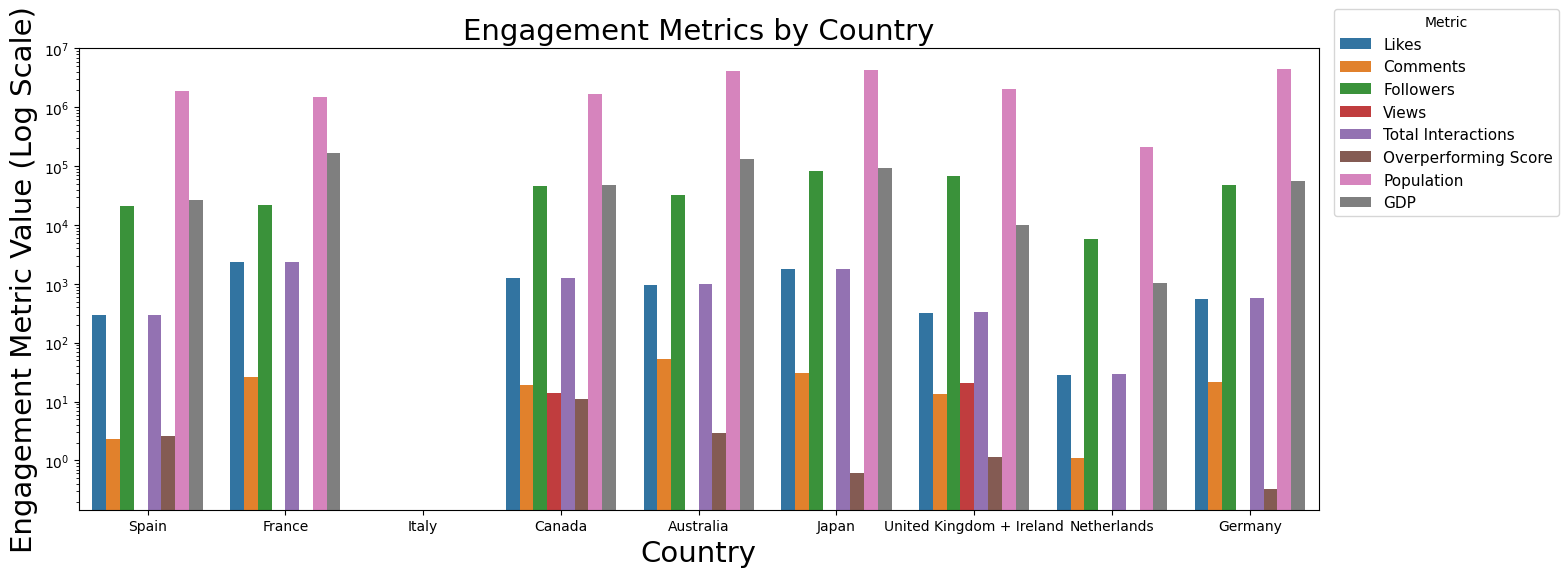} \\
            
            \bottomrule
        \end{tabularx}
    \end{adjustwidth}
\end{table}

\begin{table}[H]
    \begin{adjustwidth}{-\extralength}{0cm} 
        \begin{tabularx}{\linewidth}{m{1.5cm} X}
            \toprule
            \textbf{Outlets} & \textbf{Developing Countries} \\
            \midrule
            
            \scriptsize MCD &
            \includegraphics[width=\linewidth]{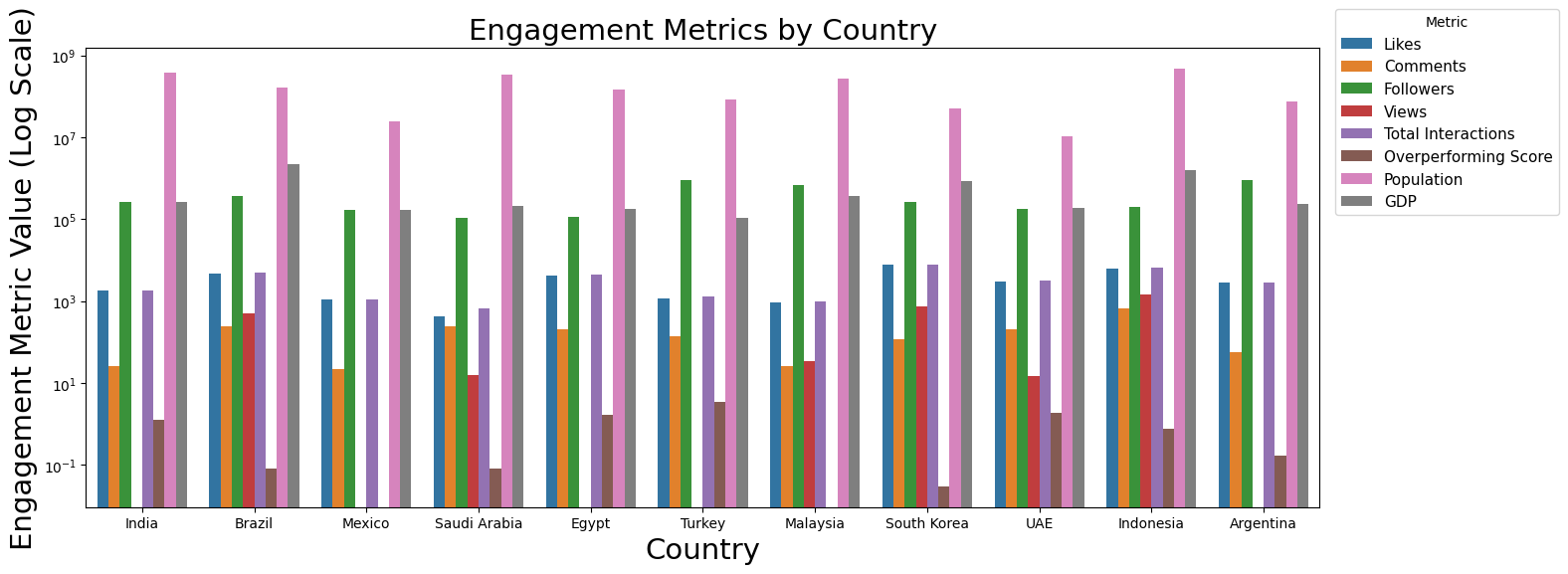} \\
            
            \addlinespace[2em]
            
            \scriptsize Starbucks &
            \includegraphics[width=\linewidth]{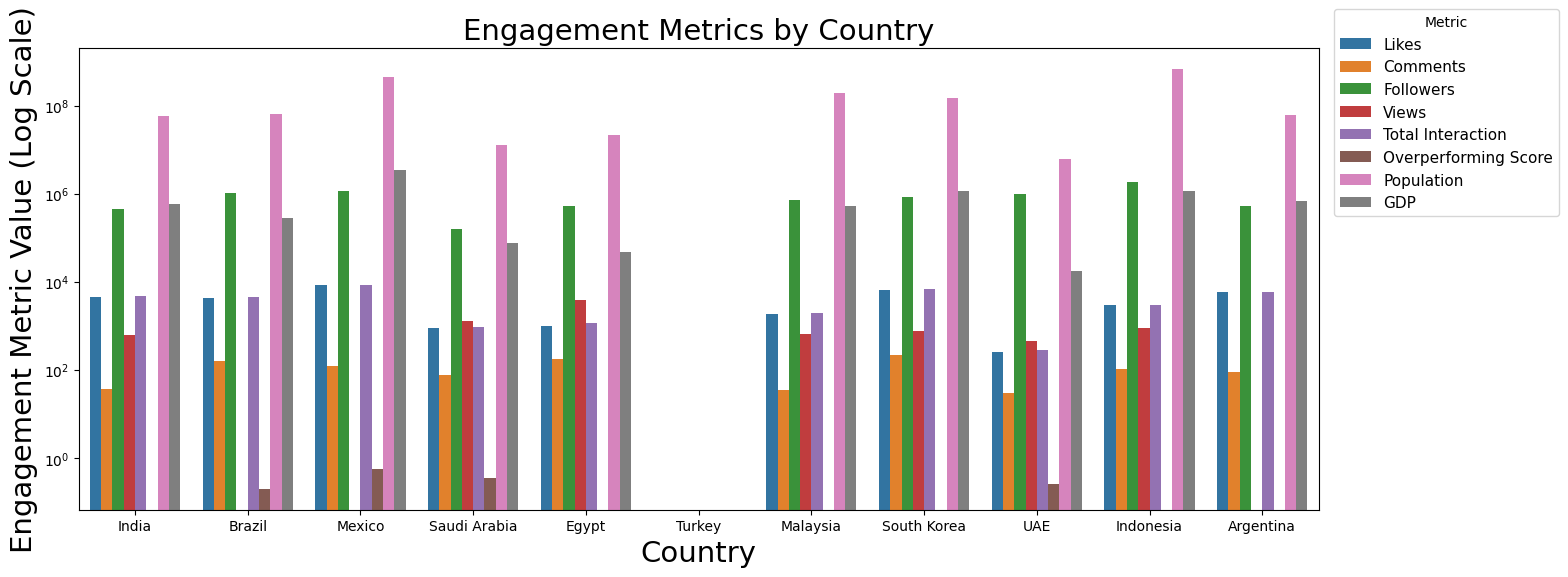} \\
            
            \bottomrule
        \end{tabularx}
    \end{adjustwidth}
\end{table}

\clearpage

\begin{table}[H]
    \begin{adjustwidth}{-\extralength}{0cm} 
        \begin{tabularx}{\linewidth}{m{1.5cm} X}
            \toprule
            \textbf{Outlets} & \textbf{Developing Countries} \\
            \midrule
            
            \scriptsize BurgerKing &
            \includegraphics[width=\linewidth]{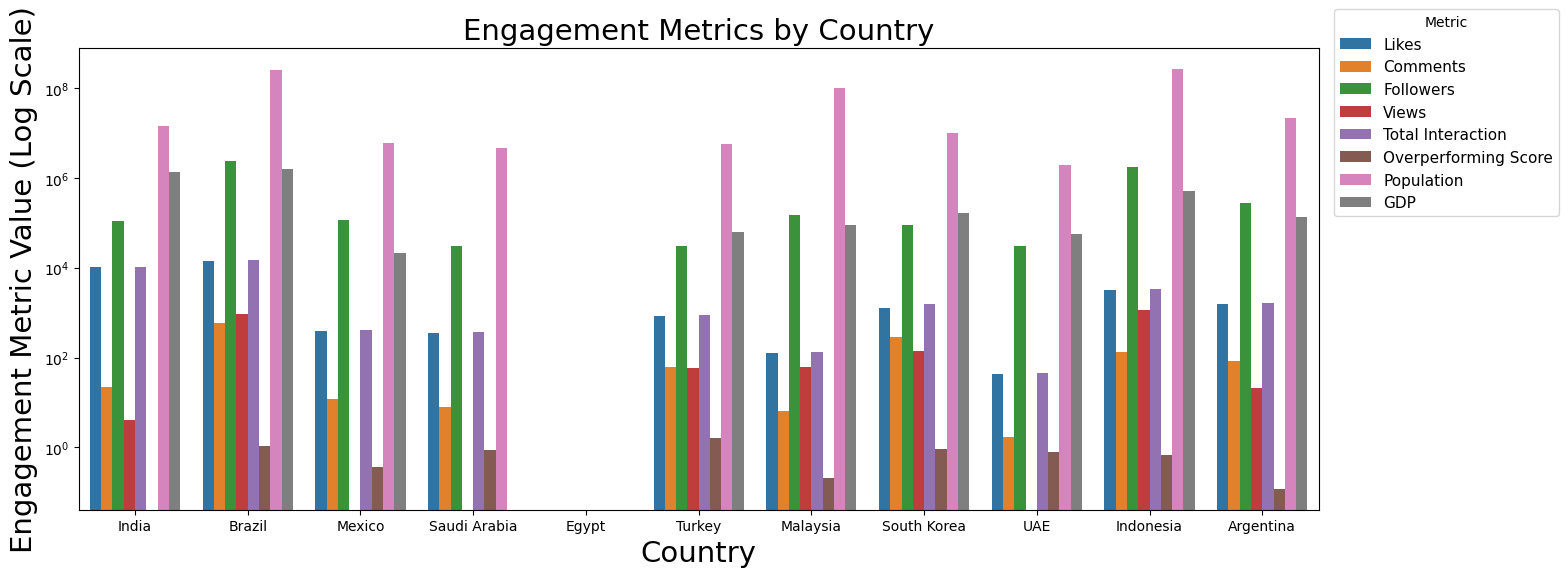} \\
            
            \addlinespace[2em]
            
            \scriptsize Dominos &
            \includegraphics[width=\linewidth]{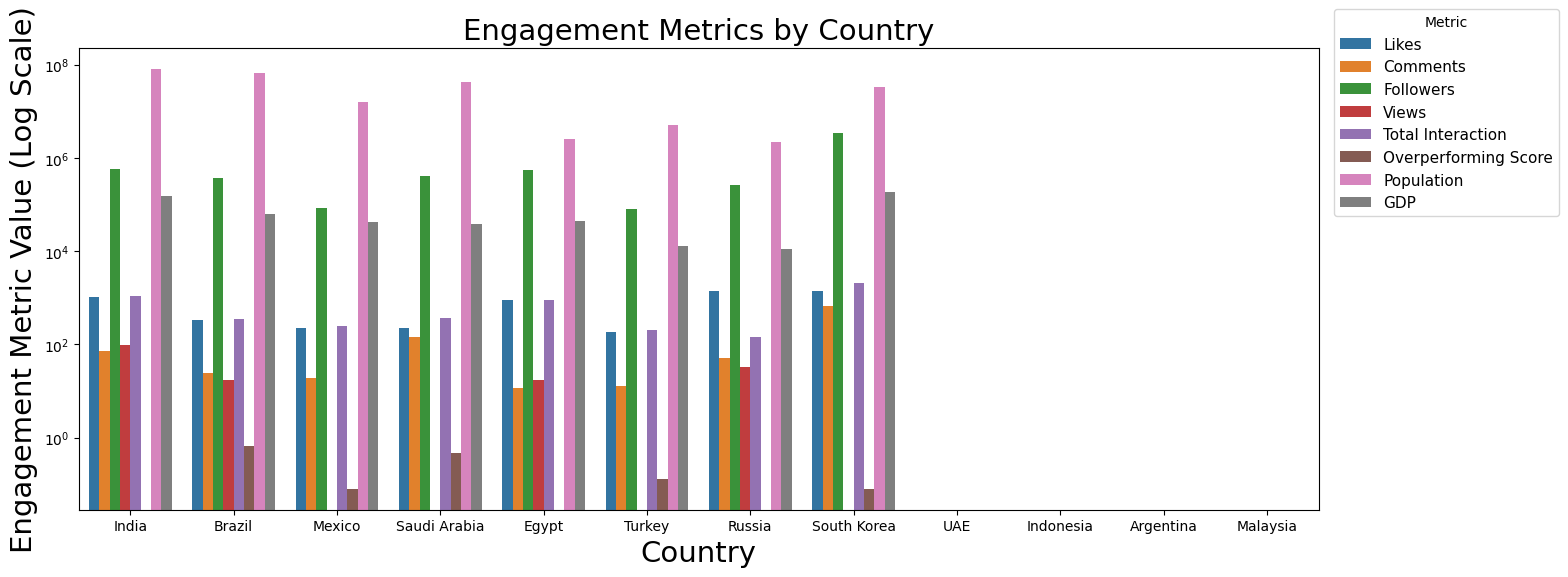} \\

            \addlinespace[2em]
            
            \scriptsize PizzaHut &
            \includegraphics[width=\linewidth]{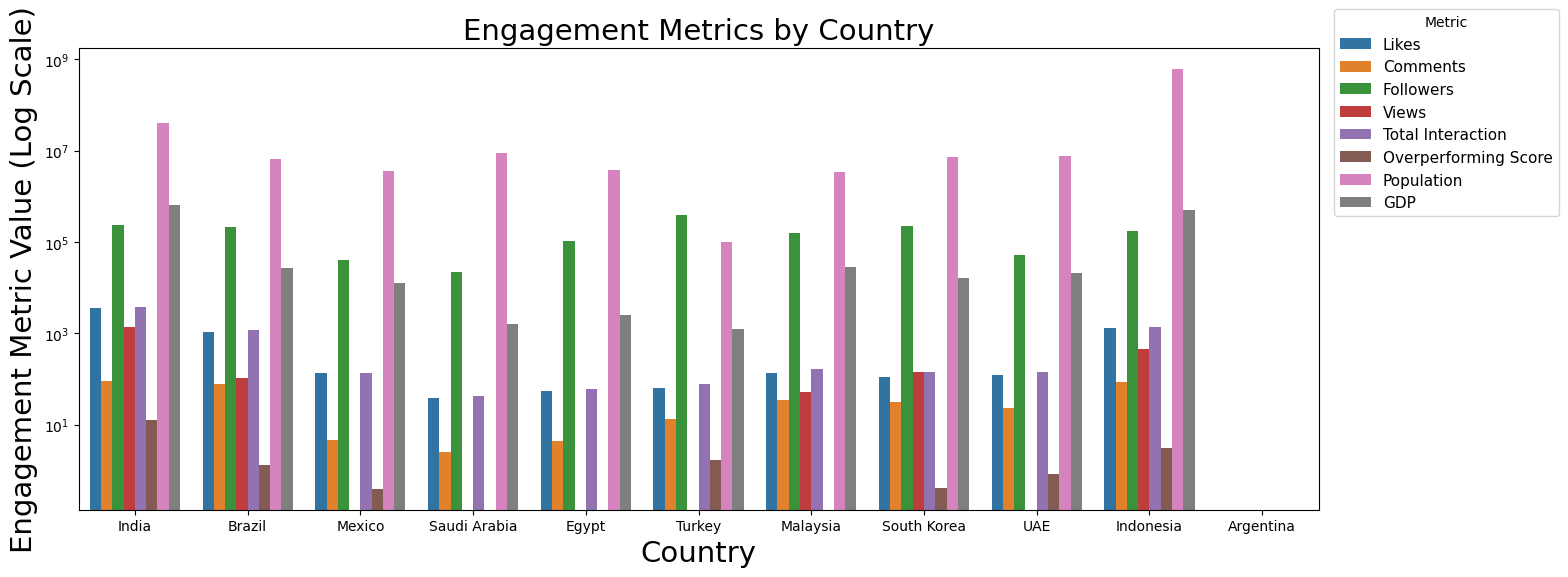} \\
            
            \bottomrule
        \end{tabularx}
    \end{adjustwidth}
\end{table}

\clearpage

\begin{table}[H]
    \begin{adjustwidth}{-\extralength}{0cm} 
        \begin{tabularx}{\linewidth}{m{1.5cm} X}
            \toprule
            \textbf{Outlets} & \textbf{Developing Countries} \\
            \midrule
            
            \scriptsize KFC &
            \includegraphics[width=\linewidth]{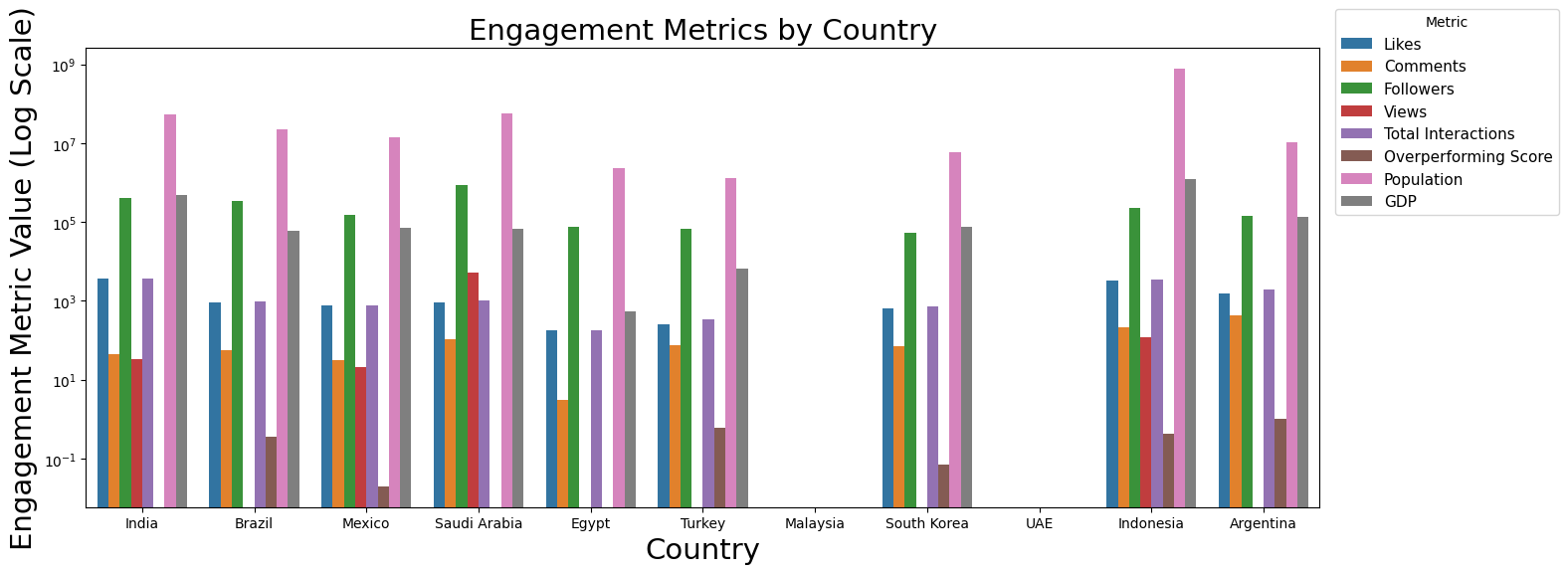} \\
            
            \addlinespace[2em]
            
            \scriptsize Subway &
            \includegraphics[width=\linewidth]{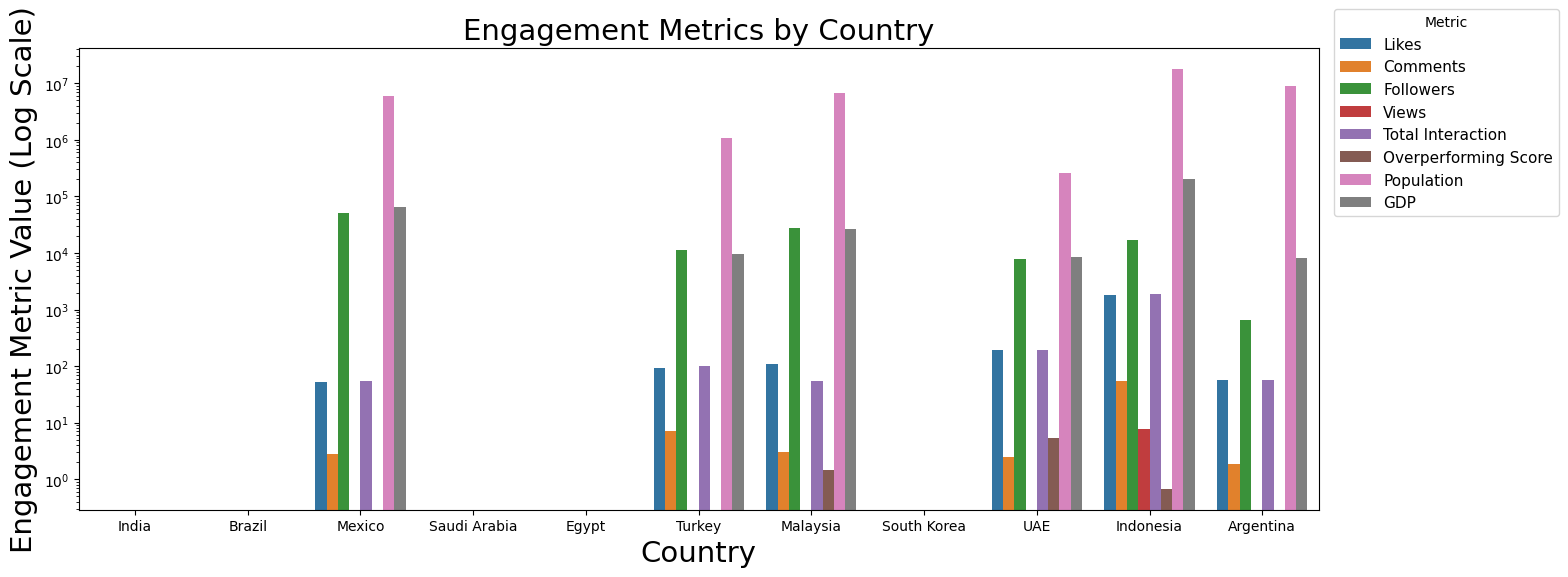} \\
            
            \bottomrule
        \end{tabularx}
    \end{adjustwidth}
\end{table}

\section*{Acknowledgments}

The authors express their sincere gratitude to our mentors, \textbf{Lynnette Hui Xian Ng} and \textbf{Swapneel Mehta}, for their invaluable guidance throughout this research. Their crucial contributions included helping formulate the core problem statement, providing critical reviews and constructive feedback on the manuscript, and offering expertise and support for the statistical analysis methods employed in the study. Their dedication greatly enhanced the clarity and rigor of this paper.

\bibliographystyle{unsrt}  
\bibliography{references}  






\end{document}